\documentclass[universe,review,accept,pdftex,moreauthors]{Definitions/mdpi} 
\firstpage{1} 
\pubvolume{9}
\issuenum{5}
\articlenumber{242}
\pubyear{2023}
\copyrightyear{2023}
\externaleditor{Academic Editor: {Daniel Cherdack} 
} 
\datereceived{19 March 2023} 
\daterevised{8 May 2023}
\dateaccepted{14 May 2023} 
\datepublished{21 May 2023} 
\hreflink{https://doi.org/10.3390/\linebreak universe9050242} 

\usepackage[flushleft]{threeparttable}
\RequirePackage{xspace}
\RequirePackage{amsmath}

\newcommand{\ttbar}{{\rm t}\bar{\rm t}}
\newcommand{\bbbar}{{\rm b}\bar{\rm b}}
\newcommand{\ttH}{{\rm t}\bar{\rm t}{\rm H}}
\newcommand{\ttV}{{\rm t}\bar{\rm t}{\rm V}}
\newcommand{\ttbb}{{\rm t}\bar{\rm t}{\rm b}\bar{\rm b}}
\newcommand{\ttb}{{\rm t}\bar{\rm t}{\rm b}}
\newcommand{\ttcc}{{\rm t}\bar{\rm t}{\rm c}\bar{\rm c}}
\newcommand{\ttc}{{\rm t}\bar{\rm t}{\rm c}}
\newcommand{\ttjj}{{\rm t}\bar{\rm t}{\rm j}\bar{\rm j}}
\newcommand{\ttcL}{{\rm t}\bar{\rm t}{\rm cl}}
\newcommand{\ttbL}{{\rm t}\bar{\rm t}{\rm bl}}
\newcommand{\ttLL}{{\rm t}\bar{\rm t}{\rm ll}}
\newcommand{\ttl}{{\rm t}\bar{\rm t}{\rm l}}
\newcommand{\sttbb}{\sigma_{\ttbb}}
\newcommand{\sttcc}{\sigma_{\ttcc}}
\newcommand{\sttjj}{\sigma_{\ttjj}}
\newcommand{\pt}{p_{\rm T}} 
\newcommand{\sqrts}{\sqrt{s}}
\newcommand{\PYTHIA} {{\textsc{pythia}}\xspace}
\newcommand{\ROOUNFOLD}{{R\textsc{oo}U\textsc{nfold}}\xspace} 
\newcommand{\MADGRAPH}{{M\textsc{ad}G\textsc{raph}}\xspace} 
\newcommand{\POWHEG}{{P\textsc{owheg}}\xspace}

\newcommand{\HERWIG}{{H\textsc{erwig}}\xspace}
\newcommand{\POWHEL}{{P\textsc{ow}H\textsc{el}}\xspace}
\newcommand{\SHERPA}{{S\textsc{herpa}}\xspace}
\newcommand{\OPENLOOPS}{{O\textsc{pen}L\textsc{oops}}\xspace}
\newcommand{\COMIX}{{C\textsc{omix}}\xspace}

\Title{{Measurements} 
 of the Cross-Section for the $\ttbar$ + Heavy-Flavor Production at the LHC}

\TitleCitation{Measurements of the Cross-Section for the $\ttbar$ + Heavy-Flavor Production at the LHC}

\Author{{Jorgen D'Hondt} 
 $^{1}$\orcidA{} and Tae Jeong Kim $^{2,}$*\orcidB{}}

\AuthorNames{Jorgen D'Hondt and Tae Jeong Kim}

\AuthorCitation{{D'Hondt, J.}
; Kim, T.J.}

\address{%
$^{1}$ \quad Inter-University Institute for High Energies (IIHE),  Vrije {Universiteit} 
 Brussel, Pleinlaan 2, \mbox{B-1050 {Brussel}
,  Belgium}; {jorgen.dhondt@vub.be} 
\\
$^{2}$ \quad Department of Physics,  Hanyang {University}
, {Seoul 04763}, Republic of Korea}

\corres{Correspondence: taekim@hanyang.ac.kr}

\abstract{
At the LHC, the process of a Higgs boson decaying into bottom or charm quarks produced in association with a pair of top quarks, $\ttH$,  allows for an empirical exploration of the heavy-flavor quark Yukawa couplings to the Higgs boson.
Accordingly, the cross-sections for the $\ttbar$ + heavy-flavor production without the appearance of the Higgs boson have been measured at the LHC in various phase spaces using data samples collected in pp collisions at $\sqrt{s}$ = 7, 8 and 13 TeV with the ATLAS and CMS experiments. Flavor ratios of cross-sections of $\ttbar$ + heavy-flavors to {$\ttbar$ + additional}
 jets processes are also measured. In this paper, the measured cross-sections and ratios are reviewed and the prospects with more data are presented. 
}


\begin{document}

\section{Introduction}

Decades of theoretical and experimental exploration of the most elementary particles and their properties yielded a detailed description of fundamental interactions, captured in a quantum field framework known as the Standard Model of particle physics. The~success of this model in describing observations over many orders of magnitude in interaction energy cannot be overestimated. 
However, despite leading to a more profound understanding, the~research field faces several problems and mysteries. Some are related to cosmological observations of dark matter in the universe and the ubiquity of matter over antimatter, some to the mathematical consistency of the model itself with respect to even the smallest variations in its parameters. Several puzzling features are related to the flavor structure of the Standard Model of particle physics, not least those present in the heavy-flavor sector. 
Through accurate measurements, we attempt to find cracks in the model where theoretical predictions may not match experimental observations. These discoveries may open new avenues to address the open problems and mysteries, either within the realm of quantum field theory or even by questioning the basic principles underlying this mathematical~framework.

After the Higgs (H) boson discovery in 2012, the~consistency check with the H boson in the standard model was one of the highest priorities at the Large Hadron Collider (LHC), especially in the heavy-flavor sector. From~analyzing the proton collision data of the LHC, the~couplings of a top quark and a bottom quark (third-generation quarks) in the standard model with the H boson were discovered in different processes~\cite{ref-atlas-ttH,ref-cms-ttH}.
However, the~confirmation that both couplings are simultaneously consistent with the predictions is only possible at the LHC by measuring the unique process of the H boson production in association with a $\ttbar$ pair ($\ttH$), where the H boson decays to a pair of bottom (b) quarks. This decay channel of the Higgs boson gives the largest signature of the $\ttbar$ pair ($\ttH$). This process alone has yet to be discovered in the data, leading to a $\ttbb$ final state. Understanding the $\ttbb$ process in proton--proton collisions without the presence of an H boson is a prerequisite to the discovery. In~addition, the~charm (c) jets in the $\ttcc$ process can also be misidentified as b jets, inducing a background in the analogy of the $\ttbb$ process. Therefore, the~measurements of cross-sections of the $\ttbar$ + heavy-flavor ($\ttbar$ + HF) process at the LHC are essential, yet challenging~objectives. 

Calculations of the inclusive production cross-section for top quark pairs with additional jets by matching matrix element generators to parton showers have been performed to next-to-leading-order (NLO) precision in quantum chromodynamics (QCD)~\cite{ref-ttjj,ref-ttjj2,ref-ttjj3,ref-ttjj4,ref-ttjj5}.
Theoretical QCD calculations of the $\ttbb$ process are available at NLO~\cite{ref-ttbb_worek,
ref-ttbb_worek1,ref-ttbb_worek2,
ref-ttbb_nlo_wishlist,ref-ttbb_lhc_1,ref-ttbb_lhc_2,ref-ttbb_lhc_3,ref-ttbb_matched} 
but they suffer from large factorization and renormalization uncertainties due to the presence of two very different scales in this process. Therefore, precise measurements can also provide a good test of the NLO QCD theory itself. 
Full NLO QCD corrections to off-shell $\ttbb$ production are available in Ref.~\cite{ref-ttbb-corr1, ref-ttbb-corr2}.
Calculations of $\ttbb$ with massive b quarks use parton density functions (PDFs) of the proton in the four flavor scheme (4FS), where b quarks are not part of the proton PDF. These matrix element level predictions of $\ttbb$ with massive b quarks are matched to parton showers~\cite{ref-ttbb_massive,ref-ttbb_powhel,ref-ttbb_nlops}.
In addition, the~associated production of $\ttbb$ with one additional jet is available as well~\cite{ref-ttbb_lightjet}. 

The cross-sections for the $\ttbar$ + HF production have been measured in various phase spaces using data samples collected in pp collisions by the ATLAS~\cite{ref-atlas} and CMS~\cite{ref-cms} experiments at $\sqrt{s}$ = 7, 8 and 13 TeV~\cite{ref-atlas-7TeV,ref-atlas-8TeV,ref-atlas-13TeV,ref-cms-ttbb-diff-8TeV,ref-cms-ttbb-dilepton-8TeV,ref-cms-ttbb-dilepton-13TeV,ref-cms-ttbb-leptonjets-13TeV,ref-cms-ttbb-hadronic-13TeV,ref-cms-ttcc-13TeV,ref-cms-ttbb-diff-13TeV}.
In order to obtain observable cross-section values, certain kinematic thresholds should be applied to the additional heavy-flavor jets. 
The interplay between the b jets from the top quark decay with the additional heavy-flavor jets is not trivial 
and, accordingly, the~definition of the signal is challenging. 
The definitions are different in each measurement and between experiments. We will discuss the definitions in Section~\ref{definition} in more detail. 
In order to achieve higher precision, 
flavor ratios of cross-sections of $\ttbar$ + HF to $\ttbar$ + additional jets processes are also measured.
The cross-section ratio measurement was originally motivated as many kinematic distributions are expected to be similar for $\ttbb$, $\ttcc$ and $\ttjj$, leading to reduced systematic uncertainties in the~ratio. 

Most measurements focus on the $\ttbb$ cross-section. The~$\ttcc$ process has been explored less due to the fact that the experimental signature of a c jet is sandwiched between that of b jets and light quark jets and gluons. With~the recent development of charm jet taggers, the~$\ttbb$ and $\ttcc$ processes can be more efficiently distinguished and the $\ttcc$ cross-section has now been measured by CMS~\cite{ref-cms-ttcc-13TeV}.   

In this experimental review, we summarize the results for the inclusive and differential cross-section measurements of $\ttbar$ + HF production at the LHC submitted to journals or available to the public before May~2023. 

\section{Definition of the \boldmath{$\ttbar$} + Heavy-Flavor~Signal}\label{definition}

\begin{figure}[H]
\includegraphics[width=6 cm]{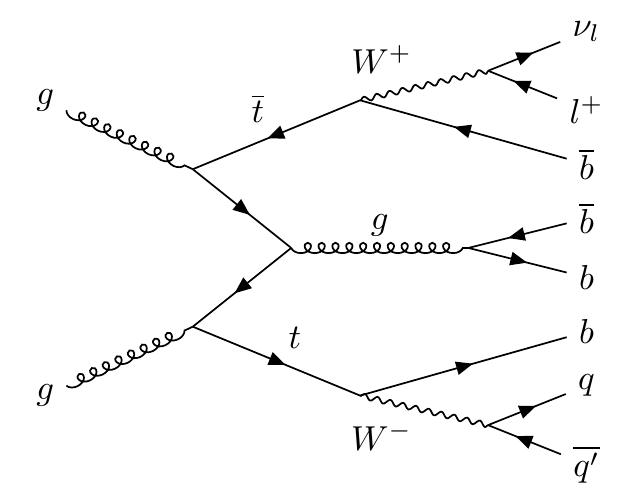}
\includegraphics[width=6 cm]{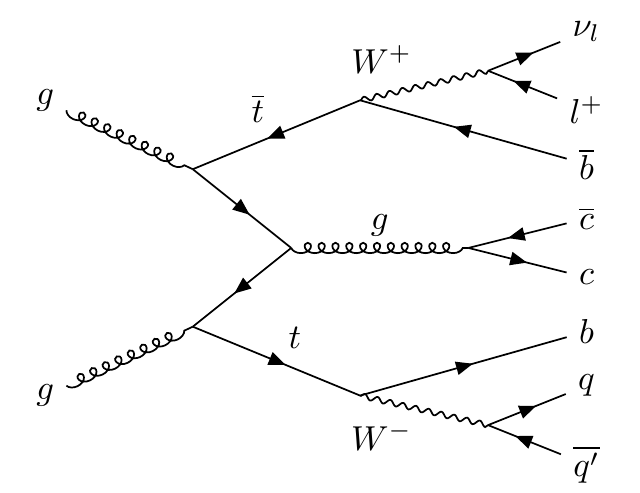}
\caption{
An example of the Feynman diagram of the $\ttbb$ and $\ttcc$ processes at the LHC in the {\mbox{lepton + jets}} 
 channel.
}\label{ttbb_feynman}
\end{figure}

The measurements of the $\ttbb$ and $\ttcc$ cross-sections are performed for both regions of the visible and the full phase space.
The resulting cross-sections at the particle level in the visible phase spaces have reduced theoretical and modeling uncertainties while the purpose of performing the measurement in the full phase space is to facilitate comparisons to theoretical calculations or measurements obtained in other decay modes. An~example of the $\ttbb$ and $\ttcc$ processes in Feynman diagrams are shown in Figure~\ref{ttbb_feynman}. Final-state particles are defined in Section~\ref{sub-object} and the processes in Section~\ref{sub-process}. 

\unskip  

\subsection{Particle-Level Object~Definition}\label{sub-object}
In the definition of the visible phase space, all generated objects such as leptons and jets are required to be within the experimentally accessible kinematic region. 
In ATLAS, the~objects are defined at the particle level which is based on the stable particles after the hadronization to reduce the dependence on the generation level information. Electrons and muons not emerging from hadron decays are considered. Furthermore, to~reach the full particle-level definition, for~all charged leptons, potential final-state photon radiation within a $\Delta R$ = 0.1 cone around the lepton is added to the four-momenta of the lepton.  
In CMS, the~electrons and muons are required to originate from a W boson at the generator level. The~electrons or muons originating from the leptonic decays of $\tau$ leptons produced in $W \to \tau \nu$ decays are included. The~procedure of adding final state photon radiation to the lepton is not performed in CMS except for the latest result in the lepton + jets channel~\cite{ref-cms-ttbb-diff-13TeV}, where the final-state photon radiation is added to the lepton at the particle level. 
The particle-level jets are defined by clustering stable particles, excluding neutrinos with the anti-kt algorithm with a distance parameter of 0.4 at a center of mass energy $\sqrt{s} = 13$ TeV and 0.5 at $\sqrt{s} = 7,~8$ TeV. 
To identify the heavy-flavor b and c particle-level jets, a~so-called ghost matching is performed. The~b- and c-hadrons are included in the jet clustering procedure after scaling their momenta to negligible values while preserving their directions.
The b and c jets are then identified by the presence of the corresponding ``ghost'' hadrons among the jet constituents. 
The approach to defining the particle-level jets is the same in ATLAS and CMS.
However, in~terms of defining the $\ttbar$ + HF quarks processes, there are subtle distinctions between different channels and experiments we will discuss in the following~section.  

\subsection{Process~Definition}\label{sub-process} 
In ATLAS, $\ttb(\bar{\rm b})$ is defined by the presence of at least three (four) particle-level b jets. Events with only three b jets can come from the case wherein one of the b jets is out of acceptance or two b jets are merged together. For~$\ttc$, in~the dilepton channel, the~number of particle-level b jets should be less than 3 and at least one c jet while in the lepton + jets channel events should contain at least two c jets as the events with exactly one c jet would come from the $W \to c\bar{s}(\bar{c}s)$ decays. If~the events with additional jets do not meet the criteria described above, the~events are grouped into a $\ttl$. In~ATLAS, only measurements in the visible phase space are available and the origin of the heavy-flavor jets is not identified. Instead, the~two b jets with the smallest $\Delta R$ separation or with the highest transverse momentum are selected. In~CMS, for~the visible phase space, the~$\ttjj$ process is defined if the event contains at least four particle-level jets including two b jets, and~the same number of leptons as required at the reconstructed level. 
The $\ttbb$ process is defined by the presence of at least four b jets regardless of their origin in the dilepton, lepton + jets channel and~hadronic channel, called ``parton-independent''. 
Additionally, in~the hadronic channel, a~parton-based definition of requiring two b jets originating from the top quark and two additional b jets is introduced.
For the $\ttbL$ process, the~event should contain only one additional b jet and at least one additional light-flavor jet or c jet.
The $\ttLL$ process is the case where 
there are no additional b or c jets, but~at least two additional light-flavor jets within the acceptance.
In the $\ttcc$ measurements, the~$\ttcc$ process is defined by the presence of at least two b jets and at least two c~jets. 

The cross-sections are measured in the visible phase space to reduce the systematic uncertainties that can be coming from the theory dependence on the acceptance. 
For the full-phase space measurements performed by CMS, the~additional b jets are required not to be from the weak decay of the $\ttbar$ system at the generator level. There is no further requirement for the decay particles from the top quarks. 
Therefore, the~measurements can be compared across different channels as well as with theory predictions.  
The cross-sections in the full phase space are obtained by taking into account the acceptance which can only be calculated based on simulations inducing an additional systematic uncertainty.  
The definitions of the signal phase space in ATLAS and CMS are summarized in Table~\ref{table:definition}.
The measured cross-sections can not be compared directly between ATLAS and CMS due to the different phase space definitions. In~the following sections, the~results from the two experiments are~reviewed.

\begin{table}[H] 
\caption{
Signal definitions at particle level in each measurement. As~a default, in~the dilepton channel, at~least two leptons are required (in ATLAS, 1 $e$ and 1 $\mu$ are required) and in the lepton + jets channel, one lepton is required exclusively. 
\label{table:definition}}
\newcolumntype{C}{>{\centering\arraybackslash}X}
\begin{threeparttable}
\begin{tabularx}{\textwidth}{CCrr}
\toprule
\textbf{Phase Space} &  ~~\textbf{Process}~~ & ~~~\textbf{ATLAS} & ~~~\textbf{CMS} \\
\midrule
\multirow{2}{*}{{Full} 
}             & $\ttbb$                       &  - &  $\geq$ 2b not from t~\cite{ref-cms-ttbb-diff-8TeV,ref-cms-ttbb-dilepton-8TeV,ref-cms-ttbb-dilepton-13TeV,ref-cms-ttbb-leptonjets-13TeV} \\
             & $\ttcc$                    &  - &  $\geq$ 2c not from t~\cite{ref-cms-ttcc-13TeV}     \\
\midrule
\multirow{5}{*}{{Visible}}             &$\ttbb$ (di-lepton)~&  $\geq$ 3(4)b~\cite{ref-atlas-7TeV,ref-atlas-8TeV,ref-atlas-13TeV} &   $\geq$ 4b~\cite{ref-cms-ttbb-diff-8TeV,ref-cms-ttbb-dilepton-8TeV,ref-cms-ttbb-dilepton-13TeV,ref-cms-ttbb-leptonjets-13TeV}   \\
             & $\ttbb$ (semi-lepton)~&   $\geq$ 5(6)j, $\geq$ 3(4)b~\cite{ref-atlas-7TeV,ref-atlas-8TeV,ref-atlas-13TeV}  &   $\geq$ 5(6)j, $\geq$ 3(4)b~\cite{ref-cms-ttbb-leptonjets-13TeV,ref-cms-ttbb-diff-13TeV} \\
             & $\ttbb$ (semi-lepton)~& - & $\geq$ 6(7)j, $\geq$ 3(4)b, $\geq$ 3l~\cite{ref-cms-ttbb-diff-13TeV} \\
             & $\ttbb$ (fully hadron)~&  - & $\geq$ 8j,  $\geq$ 4b~\cite{ref-cms-ttbb-hadronic-13TeV}  \\
             & $\ttcc$ (di-lepton)~& - & $\geq$ 2b,  $\geq$ 2c~\cite{ref-cms-ttcc-13TeV} \\
\bottomrule
\end{tabularx}
\end{threeparttable}
\end{table}
\unskip

\subsection{Monte Carlo~Simulation}\label{simulation} 

The signals of the $\ttbar$ + HF events were simulated using various Monte Carlo (MC) samples in ATLAS and CMS. Theoretical predictions are summarized in this~section.

The nominal $\ttbar$ sample was generated using the \POWHEG generator at next-to-leading-order (NLO)~\cite{ref-powheg1,ref-powheg2,ref-powheg3} at $\sqrt{s}$ = 13 TeV. The~parton shower, fragmentation and~the underlying events were simulated using $\PYTHIA$ 8.210~\cite{ref-pythia}. This sample is called \POWHEG+\PYTHIA~8~in the following. At~$\sqrt{s}$ = 8 TeV, the~events generated using the \POWHEG generator were interfaced with $\PYTHIA$ 6~\cite{ref-pythia6}. In~CMS, the~\MADGRAPH~\cite{ref-madgraph_8TeV} generator was also used as the nominal $\ttbar$ sample at $\sqrt{s}$ = 8 TeV. 
For the purpose of assessing the uncertainty due to the choice of the QCD MC model and to compare with unfolded data, alternative $\ttbar$ samples were generated. Two samples were generated using \POWHEG+\PYTHIA 8 with different renormalization and factorization scales. To~estimate the effect of the choice of parton shower and hadronization algorithms, a~$\ttbar$ sample was generated by interfacing \POWHEG with \HERWIG 7~\cite{ref-herwig1,ref-herwig2} (referred to as \POWHEG + \H 7 or as \POWHEG + \mbox{\HERWIG++} in this paper). 
The $\ttbar$ events were also generated with the \SHERPA 2.2.1 generator~\cite{ref-sherpa}, which models the zero and one additional parton process at NLO accuracy and up to four additional partons at LO accuracy. 
In addition to the samples above, a~$\ttbar$ sample was also generated using the \MADGRAPH\_aMC@NLO~\cite{ref-ttjj3}, interfaced to \PYTHIA 8. 
In CMS, the~\MADGRAPH\_aMC@NLO generator is matched to \HERWIG 6 and \PYTHIA 6 as well at $\sqrt{s}$ = 8 TeV. 
All of the $\ttbar$ samples are normalized to a cross-section calculated at next-to-next-to-leading order (NNLO)~\cite{ref-ttbar-nnlo, ref-ttbar-nnlo-top++}.

A dedicated sample of $\ttbb$ events was generated using \SHERPA+\OPENLOOPS~\cite{ref-ttbb_massive}.
The $\ttbb$ matrix elements were calculated with massive b-quarks at NLO, using the \COMIX~\cite{ref-comix} and \OPENLOOPS~\cite{ref-openloops} matrix element generators, and~merged with the \SHERPA parton shower, tuned by the authors~\cite{ref-openloops-tuned}. This sample is referred to as \SHERPA 2.2 $\ttbb$ (4FS). A~sample of $\ttbb$ events was generated using the \POWHEL~\cite{ref-ttbb_matched}, where the matrix elements were calculated at NLO with massless b-quarks and matched to the \PYTHIA 8. This sample is referred to as \POWHEL+\PYTHIA 8 $\ttbb$ (5FS). The~\POWHEL generator with massive b-quarks and matched to the \PYTHIA 8 is referred to as \POWHEL+\PYTHIA~8~$\ttbb$ (4FS). Another sample of $\ttbb$ events using the \POWHEG generator where $\ttbb$ matrix elements were calculated at NLO with mass b-quarks. The events were matched to the \PYTHIA 8. This sample is referred to as \POWHEG+\PYTHIA 8 $\ttbb$ (4FS) to distinguish it from the nominal $\ttbar$ sample described~above. 

\section{Portfolio of Cross-Section~Measurements}\label{exp}

In ATLAS, the~$\ttb$ and $\ttbb$ inclusive and differential cross-sections are measured using data 
corresponding to an integrated luminosity of 4.7 fb$^{-1}$ of proton--proton collisions 
at a center-of-mass energy of 7 TeV~\cite{ref-atlas-7TeV} and 20.3 fb$^{-1}$ at $\sqrt{s}$ = 8 TeV~\cite{ref-atlas-8TeV}. 
At $\sqrt{s}$ = 13 TeV, the~cross-sections using data corresponding to an integrated luminosity of 36.1 fb$^{-1}$ are measured in the $e\mu$ and in the lepton + jets channels~\cite{ref-atlas-13TeV}.

In CMS, the~$\ttbb$ inclusive and differential cross-sections are measured using data collected at $\sqrt{s}$ = 8 TeV~\cite{ref-cms-ttbb-dilepton-8TeV}. 
The inclusive cross-sections of the $\ttbb$ production are also measured in the dilepton channel using early data corresponding to an integrated luminosity of 2.3 fb$^{-1}$ at $\sqrt{s}$ = 13 TeV~\cite{ref-cms-ttbb-dilepton-13TeV}.
The inclusive analysis {was} updated in the dilepton channel and extended to the lepton + jets channel with data corresponding to an integrated luminosity of 35.9 fb$^{-1}$~\cite{ref-cms-ttbb-leptonjets-13TeV}. 
Measurements of the $\ttbb$ process in the hadronic channel are performed using data corresponding to an integrated luminosity of 35.9 fb$^{-1}$ at \mbox{$\sqrt{s}$ = 13 TeV}~\cite{ref-cms-ttbb-hadronic-13TeV}. 
Measurements of the $\ttcc$ production are also available at \mbox{$\sqrt{s}$ = 13 TeV}~\cite{ref-cms-ttcc-13TeV}. 
{Recently, the measurements of the inclusive and differential cross-sections were updated in the lepton + jets channel with a full Run 2 data corresponding to an integrated luminosity of 138~fb$^{-1}$~\cite{ref-cms-ttbb-diff-13TeV}.}

It is worth noting that only a small fraction of the available data has been used for all these $\sttbb$ and $\sttcc$ measurements except the lepton + jets channel.

\subsection{Inclusive Cross-Section~Measurement}\label{exp-inclusive}

In ATLAS, at~$\sqrt{s}$ = 13 TeV, the~cross-section measurements were performed in the $e\mu$ channel within the at least three b jet visible phase space and in lepton + jets channels within the at least four b jet visible phase space. 
To extract the $\ttbar$ + HF number of events, in~both channels, a~binned maximum likelihood fit is used on observables discriminating between signal and background. A~combined template is created from the sum of all backgrounds. Three templates of $\ttb$, $\ttc$ and $\ttl$ events are created from all of $\ttbar$, $\ttbar$ in association with a vector boson ($\ttV$) and $\ttH$ simulations as those samples contain the signal process. In~the $e\mu$ channel, $\ttc$ and $\ttl$ are merged together to fit to the distribution of the third highest b-tagging discriminant among the reconstructed jets in the event. The~scale factors obtained from the fit are 1.33 $\pm$ 0.06 for the number of $\ttb$ events and 1.05 $\pm$ 0.04 for the number of the combined $\ttc$ + $\ttl$ events. 
In the lepton + jets channel, all three templates are used to fit to the 2D histograms of the third and fourth b-tagging discriminant. The~best fit values are 1.11 $\pm$ 0.2 for the number of $\ttb$ events, 1.59 $\pm$ 0.06 for the number of $\ttc$ events and \mbox{0.962 $\pm$ 0.003} for the number of $\ttl$ events.  The~measured cross-section values for $\ttb$ for both channels are compatible with each~other.

\begin{figure}[H]
\includegraphics[width=12 cm]{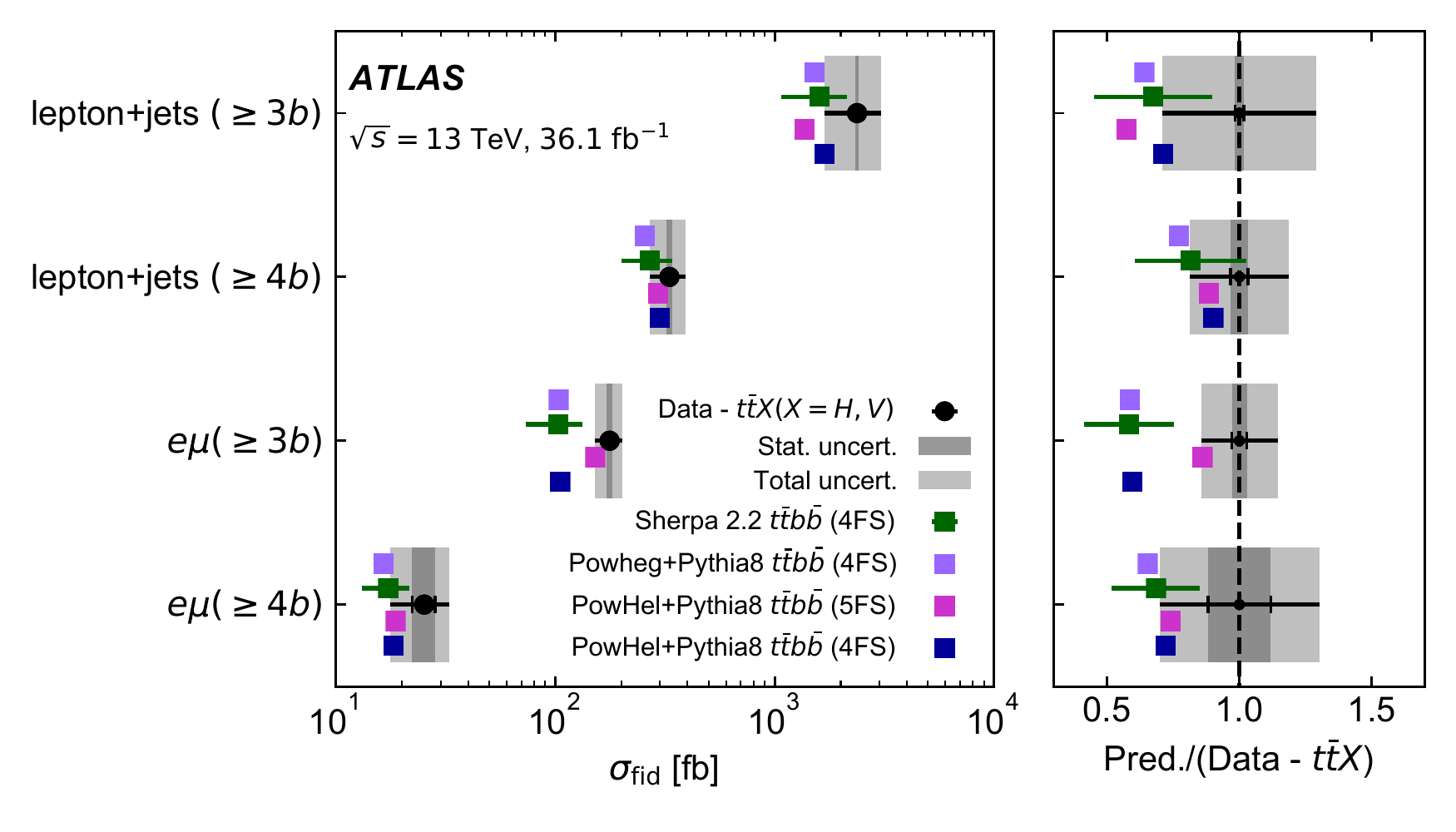}
\caption{The visible phase space cross-sections measured by ATLAS compared with $\ttbb$ predictions obtained using \SHERPA 2.2, \POWHEG+\PYTHIA 8 and \POWHEL+\PYTHIA 8 $\ttbb$.
The $\ttH$ and $\ttV$ processes are subtracted from the measurement to facilitate the comparison with theory~{\cite{ref-atlas-13TeV}}
.
}\label{atlas_inclusive_13TeV}
\end{figure}

To facilitate the comparison with the theory $\ttbb$ cross-section, the~$\ttH$ and $\ttV$ processes are also subtracted from the measured cross-section. The~measured inclusive cross-sections are shown in Figure~\ref{atlas_inclusive_13TeV}. 
All of the inclusive cross-sections measured at \mbox{$\sqrt{s}$ = 13 TeV} in the visible phase space by the ATLAS experiment are summarized in Table~\ref{table:atlas_visible}.
The cross-section measurement for the $\geq$three b jet phase space in the $e\mu$ channel has an uncertainty of 13\%, which is the most precise measurement.   
The uncertainties are dominated by systematic uncertainties mainly from the $\ttbar$ modeling and b-tagging, as well as~the jet energy~scale. 

\unskip   

\begin{table}[H] 
\caption{
Measured $\ttbb$ cross-sections compared with predictions 
from \SHERPA 2.2 in the visible phase space at $\sqrt{s}$ = 13 TeV by ATLAS. 
The $\ttH$ and $\ttV$ contributions are subtracted from the measurement.
The statistical and dominating systematical uncertainties on the measurements are presented.  
\label{table:atlas_visible}}
\newcolumntype{C}{>{\centering\arraybackslash}X}
\begin{threeparttable}
\begin{tabularx}{\textwidth}{lCCC}
\toprule
\textbf{Channel}	   & \textbf{Measurements (\textbf{pb}) }            & \textbf{Predictions (\textbf{pb})} 	& \textbf{Phase Space}\\
\midrule
$\ttbb$   \\ 
($e\mu$)~\cite{ref-atlas-13TeV}        &  177 $\pm$ 5 $\pm$ 24     & 103 $\pm$ 30	     &  $\geq$3b \\
($e\mu$)~\cite{ref-atlas-13TeV}         &  25 
 $\pm$ 3 $\pm$ 7      & 17.3 $\pm$ 4.2	 & $\geq$4b \\
(lepton + jets)~\cite{ref-atlas-13TeV}    & ~ 2370 $\pm$ 40 $\pm$ 690  & 1600 $\pm$ 530       & $\geq$ 5j, $\geq$3b \\
(lepton + jets)~\cite{ref-atlas-13TeV}    & ~ 331 $\pm$ 11 $\pm$ 61    & 270 $\pm$ 70         & $\geq$ 6j, $\geq$4b  \\
\bottomrule
\end{tabularx}
\end{threeparttable}
\end{table}

\begin{figure}[H]
\includegraphics[width=12 cm]{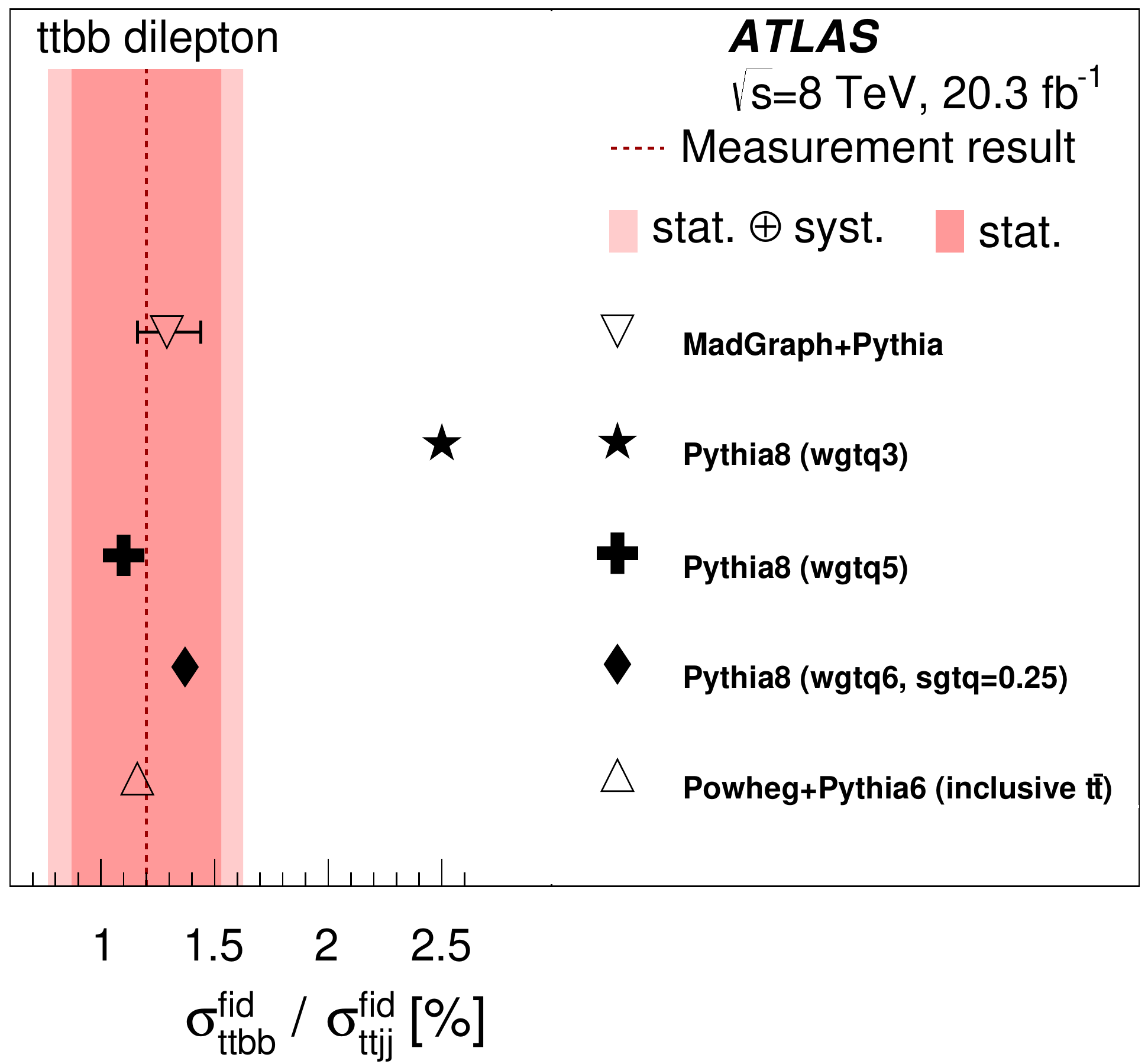}
\caption{{Measurement} 
 of the ratio between the $\ttbb$ and $\ttjj$ visible cross-sections at $\sqrt{s}$=8 TeV by ATLAS~\cite{ref-atlas-8TeV}.
}\label{atlas_ratio_8TeV}
\end{figure}

The ratio measurement of the cross-sections of $\ttbb$ to $\ttjj$ production is also available using data collected at $\sqrt{s}$ = 8 TeV~\cite{ref-atlas-8TeV}. The~ratio measurement is motivated to reduce the systematic uncertainties and the result is compared with predictions in Figure~\ref{atlas_ratio_8TeV}.

\begin{figure}[H]
\includegraphics[width=12 cm]{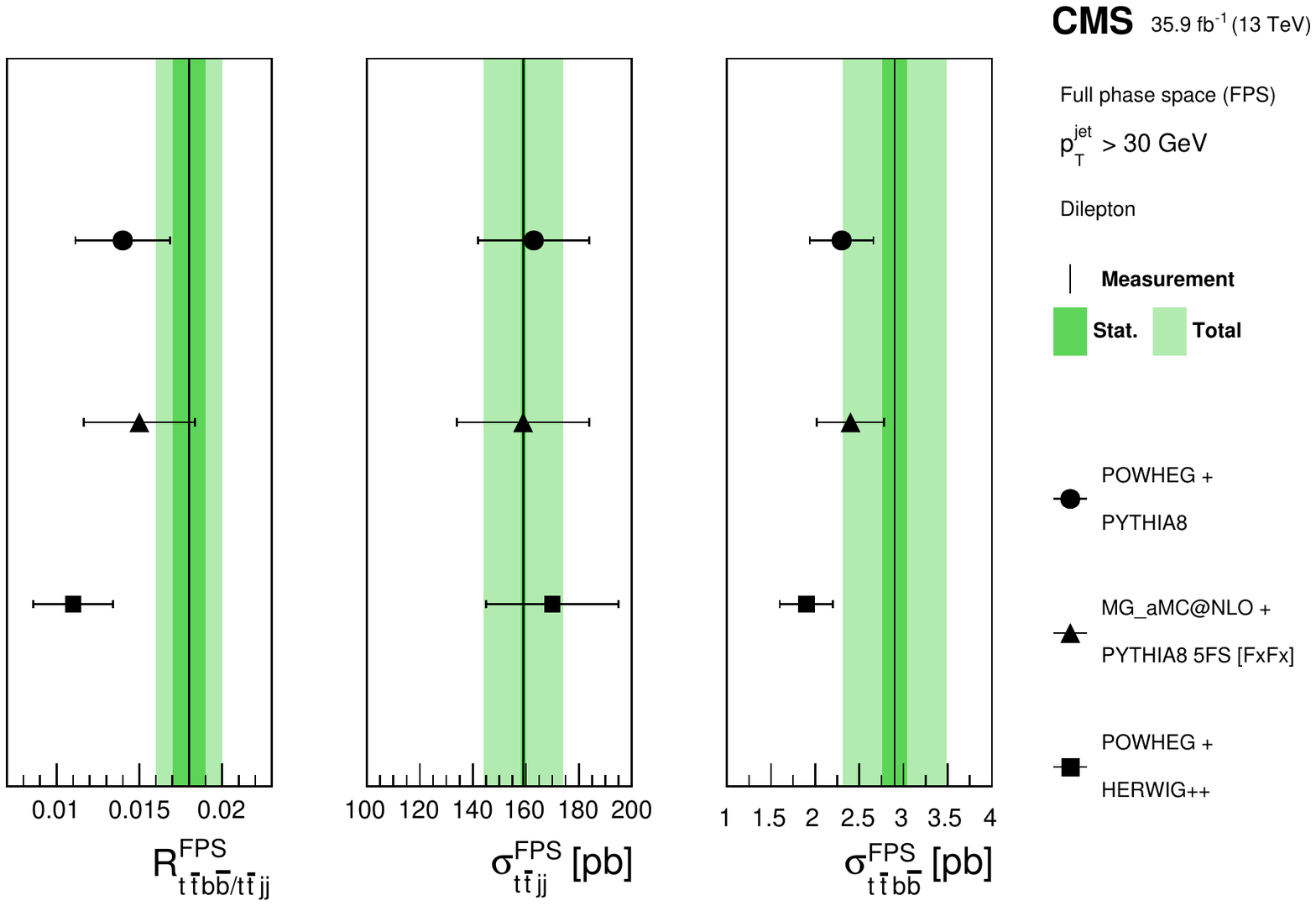}\\
\includegraphics[width=12 cm]{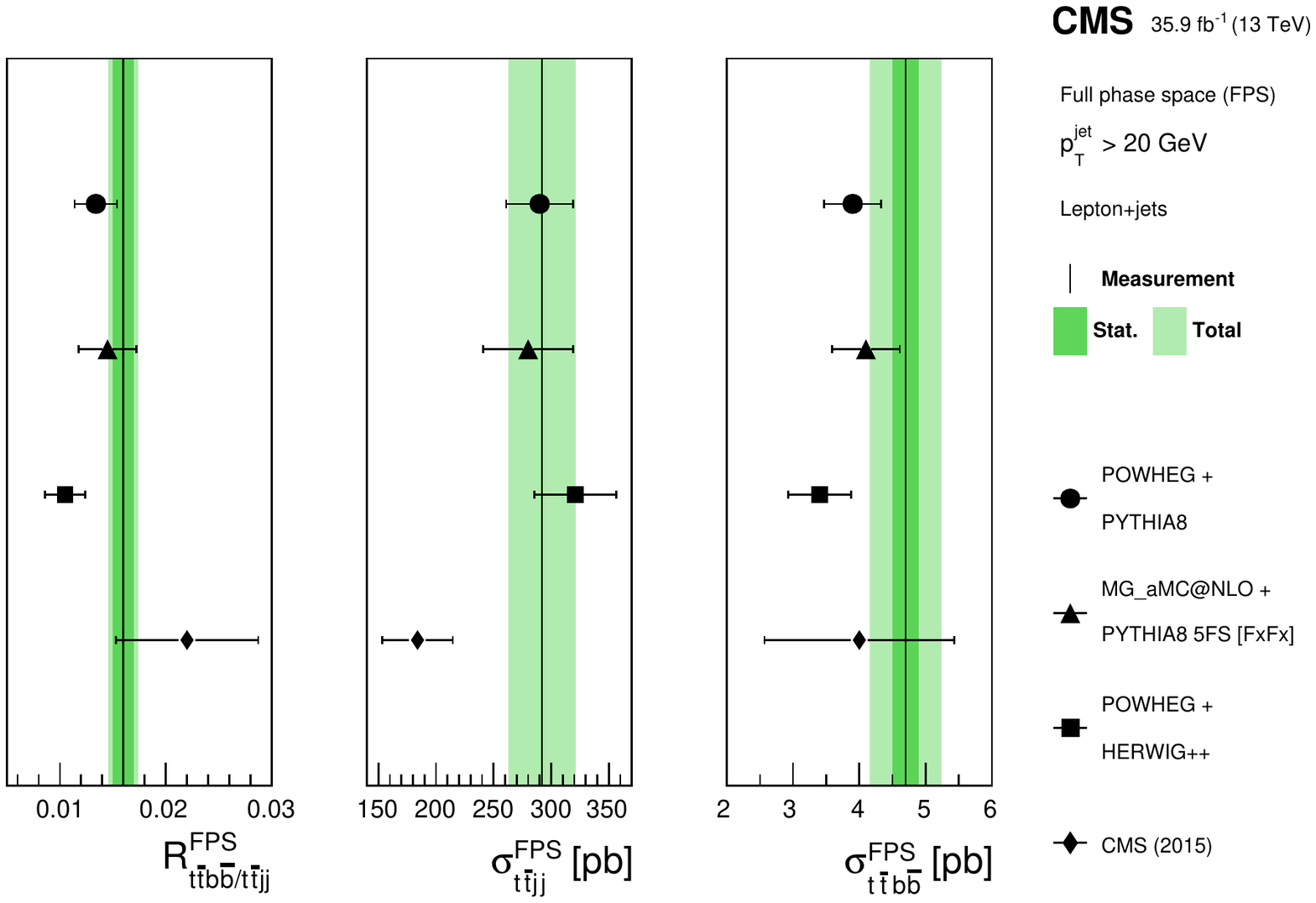}
\caption{{The} 
 measured $\ttbb$ cross-sections in the full phase space in the dilepton channel (\textbf{upper}) and the lepton + jet channel (\textbf{lower})
from the CMS experiment.
The dark (light) shaded bands show the statistical (total) uncertainties on the measured values~\cite{ref-cms-ttbb-leptonjets-13TeV}.
\label{CMS-TOP-18-002}}
\end{figure}

In CMS, the~inclusive $\ttbb$ cross-sections are measured in the different phase spaces of the dilepton, lepton + jets and hadronic channels using data collected at $\sqrts$ = 13 TeV by CMS. In~the dilepton channel, measurements at $\sqrt{s}$ = 8 TeV are also available. In~the dilepton channel, the~final state consists of two reconstructed leptons and at least four reconstructed b jets. With~these two leptons, the~dominating Z + jets background is estimated from data using control samples enriched in Z boson events. Among~the at least four b jets, the~first and the second jets in decreasing order of the b tagging discriminator tend to be the b jets from the top quark. 
Therefore, jets with the third and fourth largest b tagging discriminator are considered as the additional b jets. 
Using the two-dimensional distribution of these discriminators of two determined additional jets, 
the number of $\ttbb$ events is extracted. 
Together with the ratio $\sttbb/\sttjj$, the~cross-sections $\sttbb$ and $\sttjj$ are measured in the visible phase space. 
For the purpose of comparing the measurements with the theoretical prediction and with measurements in the other decay modes, the~cross-sections in the full phase space are obtained by taking into account the acceptance, $\sigma_{full} = \sigma_{visible} /\mathcal{A}$, where $\mathcal{A}$ is the acceptance, defined as the number of events in the corresponding visible phase space divided by the number of events in the full phase space. The~results for the full phase space are shown in Figure~\ref{CMS-TOP-18-002} (upper).

In the lepton + jets channel, the~measurement was conducted with data corresponding to an integrated luminosity of 35.9 fb$^{-1}$ at $\sqrt{s}$ = 13 TeV in CMS. In~this channel, the~identification of the origin of the jets is challenging because the final state with at least six jets including four b jets leads to ambiguities in the jet assignment. Moreover, the~heavy-flavor jet can also originate from the W boson decay. In order to address this, the~kinematic reconstruction method is used to identify the additional b jets. The~algorithm assigns a $\chi^2$ value according to the goodness of fit of each jet permutation to meet certain kinematic constraints. 
The solution selected is the one with the lowest $\chi^2$ value. 
Once a jet topology is selected, the~additional jets in the event are arranged in decreasing order of their b tagging discriminant value. 
Then, similar to the dilepton channel, only the information from two additional jets with the highest b tagging discriminant value is used to extract the $\ttbb$ cross-section. The~results for the ratio $\sttbb/\sttjj$, $\sttbb$ and $\sttjj$ are presented for both the visible phase space and the full phase space (see Figure~\ref{CMS-TOP-18-002}). Recently, the measurement in the lepton + jets channel was updated with a full Run 2 data corresponding to an integrated luminosity of 138 fb$^{-1}$~\cite{ref-cms-ttbb-diff-13TeV}.
In this analysis, the~cross-sections in four different visible phase spaces are measured extensively in four different phase spaces. The~final states of each phase space are shown in Table~\ref{table:definition}. 
For the phase spaces with the requirement of three additional light jets, it is motivated for the study of additional QCD radiation in $\ttb$ or $\ttbb$ events as these have been shown to be sensitive to the modeling of $\ttbb$ production.
The measured cross-sections in all phase spaces are larger than the predictions from the \POWHEG+\PYTHIA8. All other predicted values in each phase space are available in Ref.~\cite{ref-cms-ttbb-diff-13TeV}.

\begin{figure}[H]
\includegraphics[width=10.5 cm]{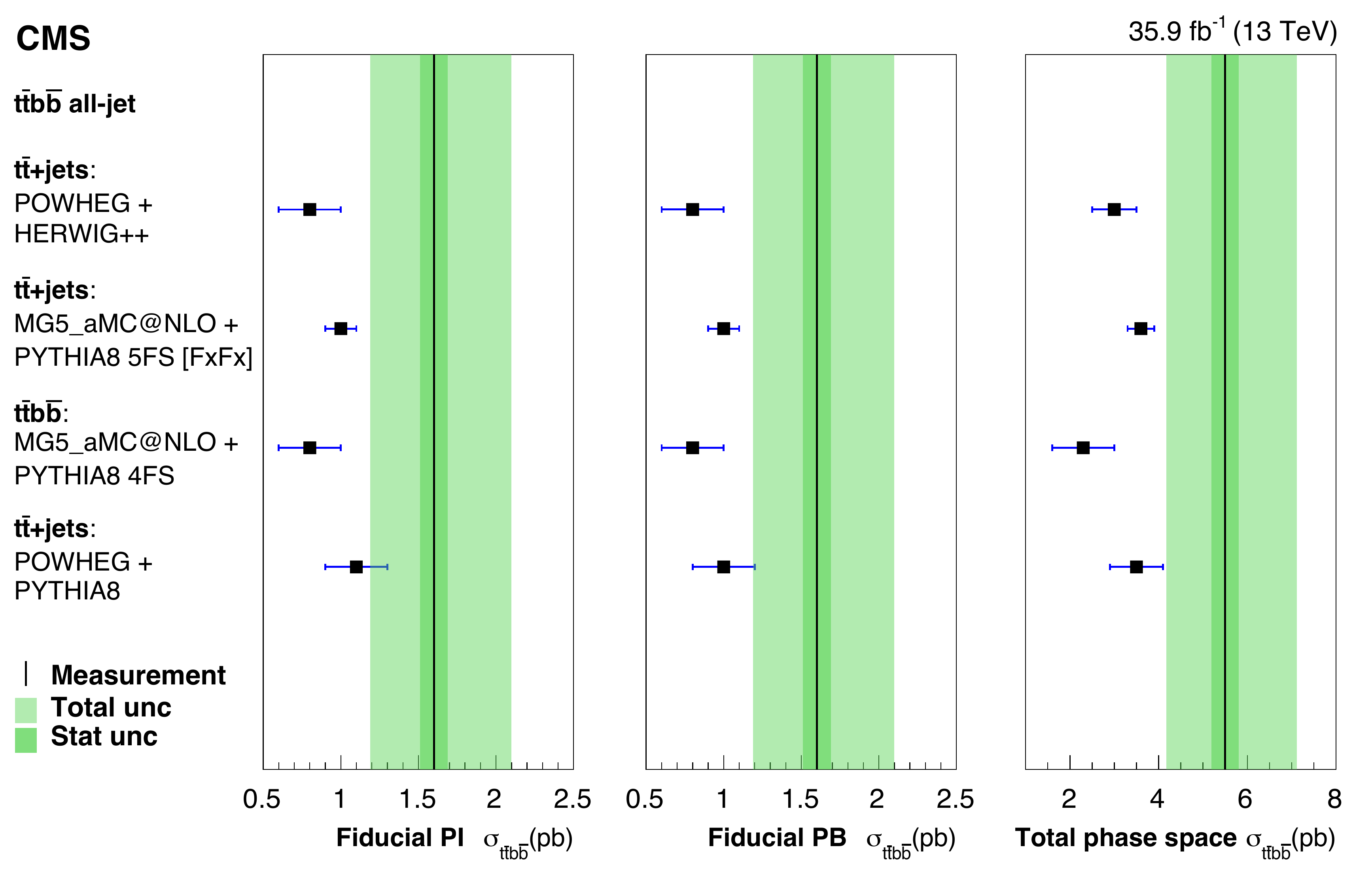}
\caption{{The} 
 measured $\ttbb$ cross-sections in the hadronic channel in the parton-independent (\textbf{left}), parton-based (\textbf{center}) and full phase space (\textbf{right}) from the CMS experiment.
The dark (light) shaded bands show the statistical (total) uncertainties on the measured values~\cite{ref-cms-ttbb-hadronic-13TeV}.
\label{CMS-TOP-18-011}}
\end{figure}

In the hadronic channel, the~multi-jet process is the main background. To~remove the multi-jet events, the~quark--gluon discriminant was used. The~unsupervised learning algorithm was also further used to maximize the contribution of $\ttbb$ events.
The measured cross-sections follow two definitions of the $\ttbb$ events in the fiducial phase space. One is based exclusively on stable generated particles after hadronization (parton-independent). This definition facilitates comparisons with predictions from event generators. The~other uses parton-level information after radiation emission (parton-based). 
This definition is closer to the approach taken by searches for $\ttH$ production to define the contribution from the $\ttbb$ process. 
To address the large combinatorial ambiguity in identifying the additional jets in the events, a~boosted decision tree (BDT) was used.
The cross-section is also reported for the total phase space by correcting the parton-based fiducial cross-section by the experimental acceptance. 
The results are presented in Figure \ref{CMS-TOP-18-011}.

\vspace{-3pt}
\unskip

The cross-section of a top quark pair production with an additional pair of c jets has been measured for the first time by CMS.
This measurement is challenging as the experimental signature of a b jet is very similar to that of a c jet. 
Two additional jets are selected using a deep neural network classifier. 
To separate the $\ttcc$, $\ttbb$ and~$\ttLL$ events, a~NN is trained  
using charm jet tagging information of the first and second additional jets, and~kinematic variables such as the angular separation $\Delta R$ between two additional jets, as well as~the NN score for the best jet permutation. This NN predicts the probabilities for five output classes of $\ttcc$, $\ttcL$, $\ttbb$, $\ttbL$ and $\ttLL$. Two discriminators are derived as follows.
\begin{equation}
\begin{split}
    \Delta^c_b = \frac{P(\ttcc)}{P(\ttcc) + P(\ttbb)}, \\
    \Delta^c_L = \frac{P(\ttcc)}{P(\ttcc) + P(\ttLL)}.
\end{split}
\end{equation}

\begin{figure}[H]
\includegraphics[width=6.75 cm]{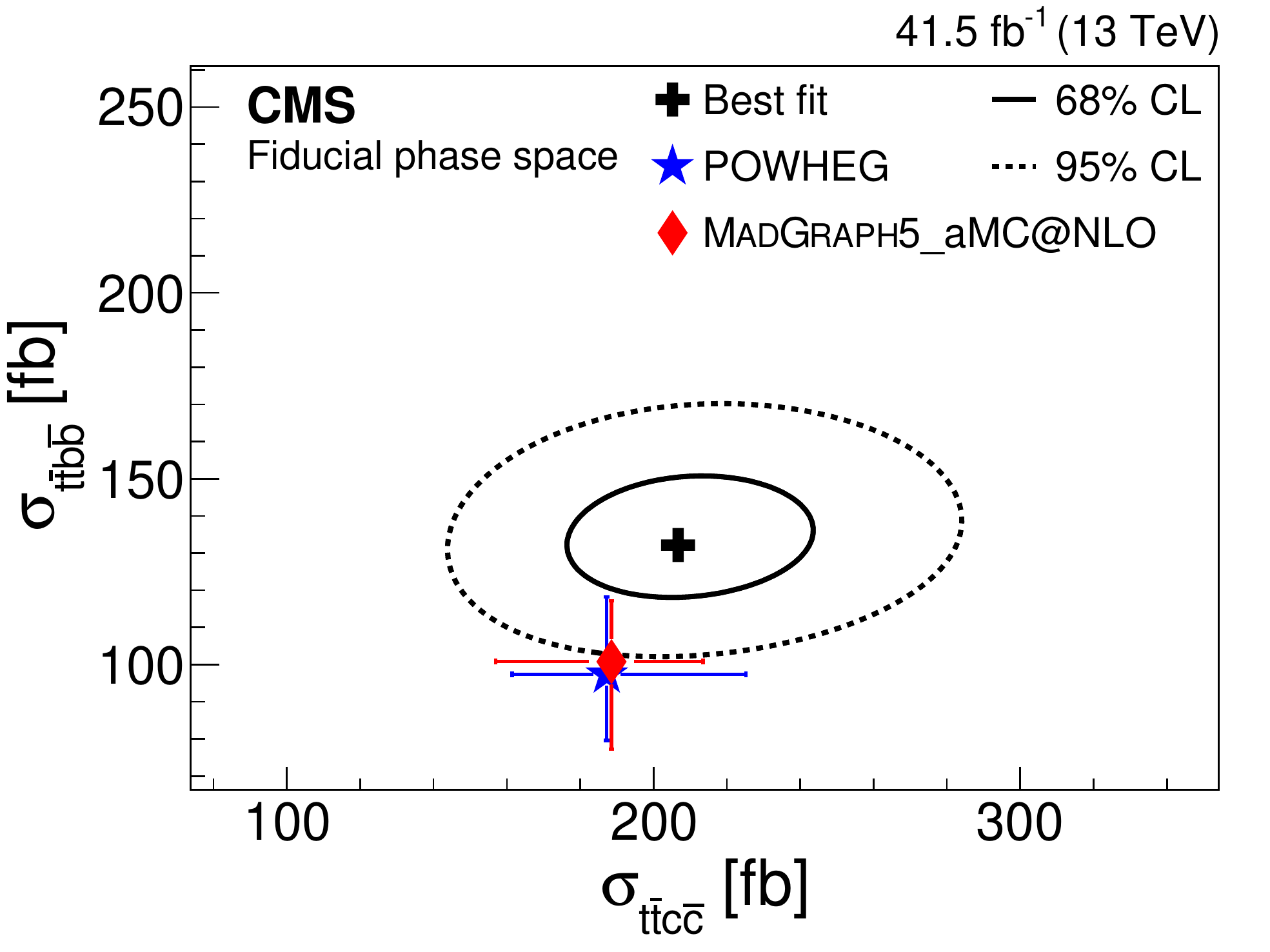}
\includegraphics[width=6.75 cm]{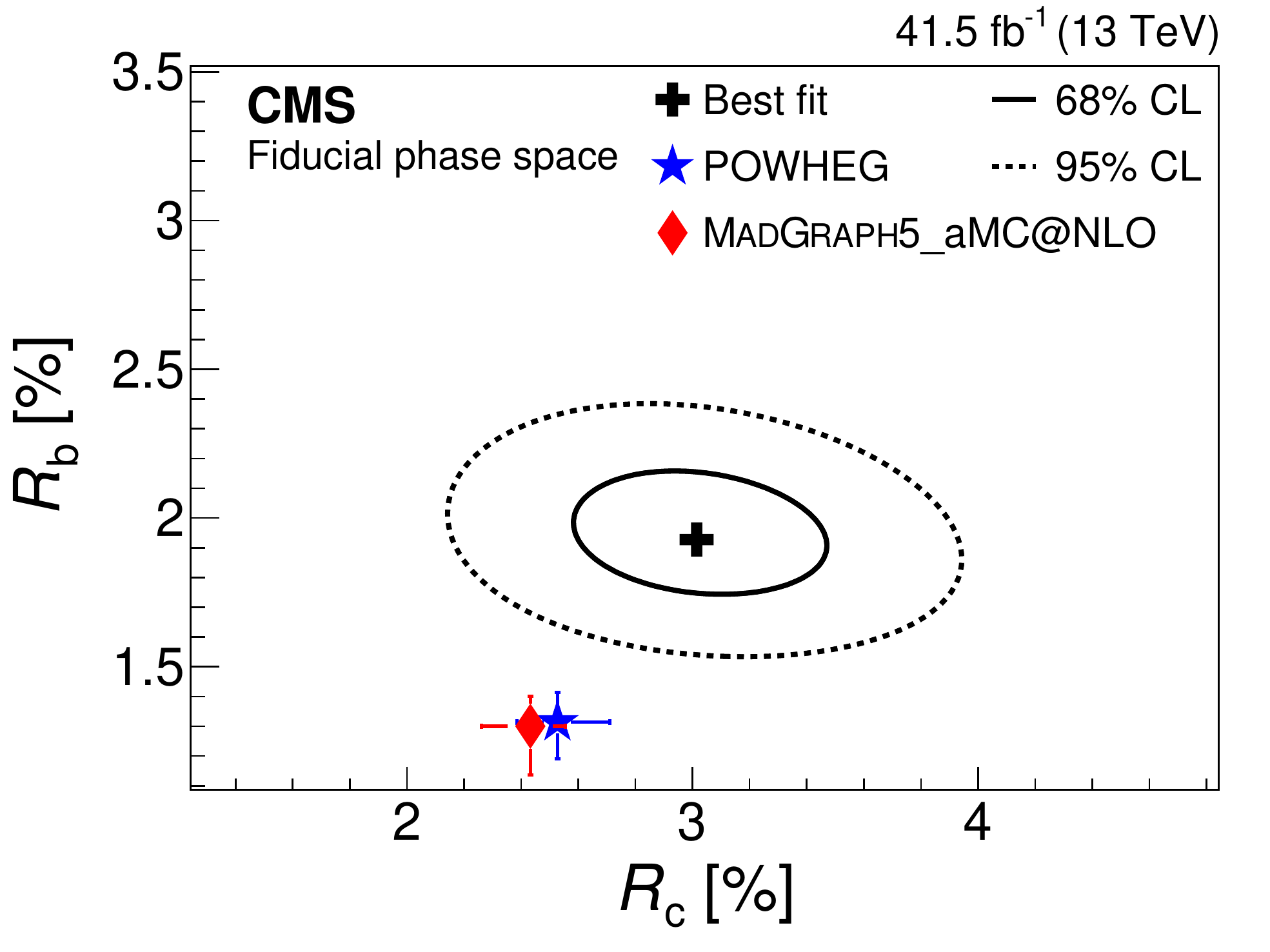}
\caption{
{Results} 
 of the $\ttbb$ versus $\ttcc$ cross-section measured by CMS in the fiducial phase space, and~their ratios to the inclusive $\ttbar$ + two jets cross-section \cite{ref-cms-ttcc-13TeV}.
\label{CMS_ttcc}}
\end{figure}

The $\ttcc$, $\ttbb$ and~$\ttLL$ cross-sections are extracted from a fit to the two-dimensional distribution of these discriminators. 
The ratios $R_b$ and $R_c$ of, respectively, the~measured $\sttbb$ and $\sttcc$ cross-sections with respect to the inclusive $\ttbar$ + two jets cross-section were also measured. The~results are compared to theoretical predictions of either the \POWHEG or \MADGRAPH5\_aMC@NLO generators as shown in Figure~\ref{CMS_ttcc}.

\vspace{-3pt}

\begin{figure}[H]
\includegraphics[width=11 cm]{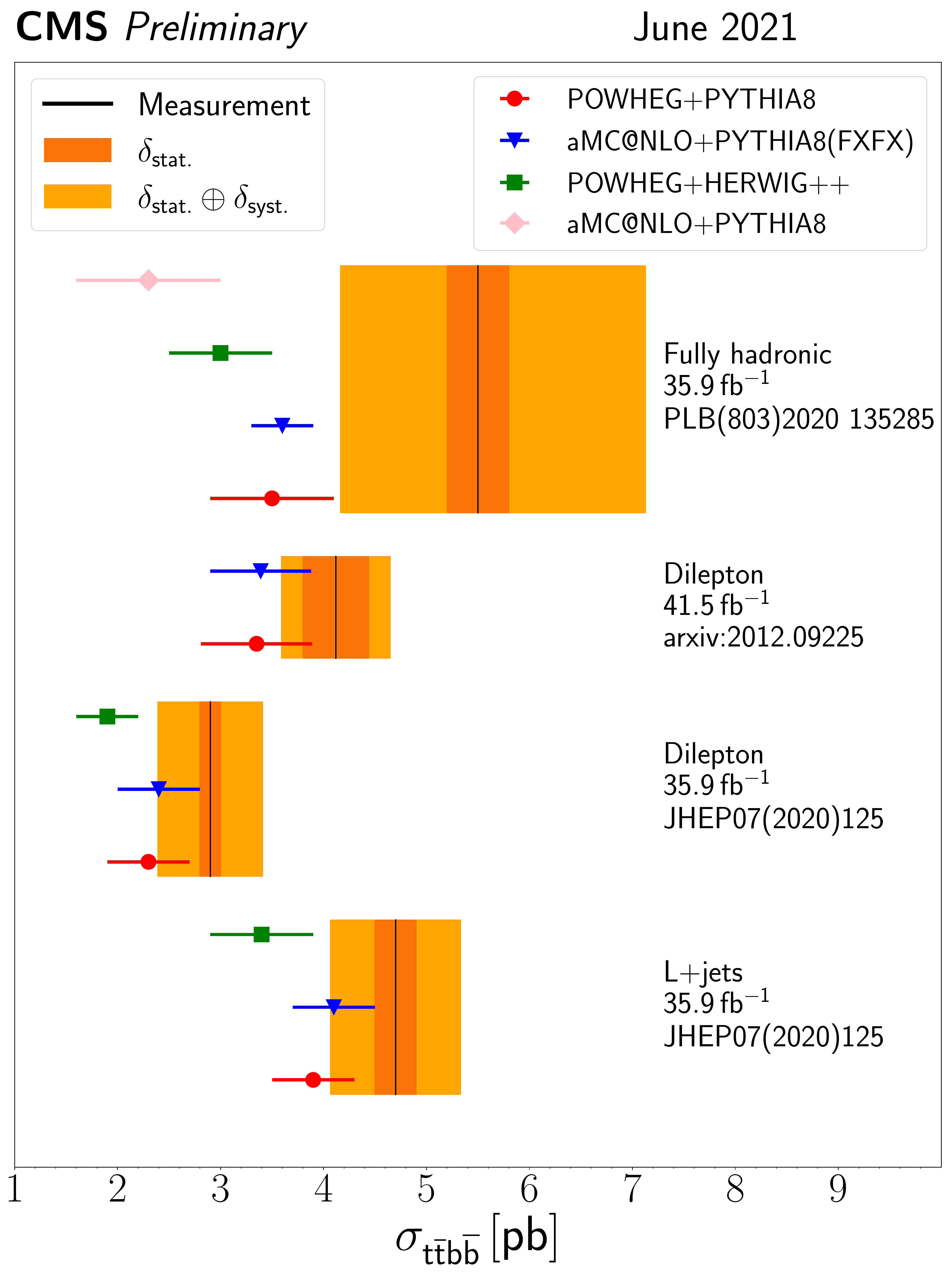}
\caption{
The $\ttbb$ cross-sections in the various channels for the full phase space measured by CMS~\cite{cms_top_plots}.
}\label{cms_ttbb_summary}
\end{figure}

All of the inclusive cross-sections measured in the visible phase space by the CMS experiment are summarized in Tables~\ref{table:visible_cross} and~\ref{table:visible_ratio}, and~for the full phase space in Tables~\ref{table:full_cross} and~\ref{table:full_ratio}. 
Figure~\ref{cms_ttbb_summary} also shows the comparison between the measured values in the full phase space and various theoretical predictions in~CMS. 

\begin{table}[H] 
\caption{Measured and predicted cross-sections in the visible phase space at $\sqrt{s}$ = 13 TeV by CMS. In~the hadronic channel, the~parton-based cross-section is shown. The~predictions of \POWHEG+\PYTHIA8 are shown. 
\label{table:visible_cross}}
\newcolumntype{C}{>{\centering\arraybackslash}X}
\begin{threeparttable}
\begin{tabularx}{\textwidth}{lCCC}
\toprule
\textbf{Channel}	   & \textbf{Measurements (\textbf{pb}) }            & \textbf{Predictions (\textbf{pb})} 	& \textbf{b jet Requirement}\\
\midrule
$\ttbb$\\
dilepton~\cite{ref-cms-ttbb-dilepton-13TeV}      & ~~0.088 $\pm$ 0.012 $\pm$ 0.029 & 0.070 $\pm$ 0.009  &  $\pt$ $>$ 20 GeV, $|\eta|$ $<$ 2.5   \\
dilepton~\cite{ref-cms-ttbb-leptonjets-13TeV}    & ~~0.040 $\pm$ 0.002 $\pm$ 0.005 & 0.032 $\pm$ 0.004  & $\pt$ $>$ 30 GeV, $|\eta|$ $<$ 2.4    \\
lepton + jets   &  &  &  \\ 
~~$\geq$ 6j $\geq$ 4b~\cite{ref-cms-ttbb-leptonjets-13TeV} & ~~0.62 ~$\pm$~ 0.03 ~$\pm$~ 0.07 & 0.52 $\pm$ 0.06  & $\pt$ $>$ 20 GeV, $|\eta|$ $<$ 2.5  \\
~~$\geq$ 5j, $\geq$ 3b~\cite{ref-cms-ttbb-diff-13TeV} & ~~2.368 $\pm$ 0.142 $\pm$ 0.014 & 1.791 & $\pt$ $>$ 25 GeV, $|\eta|$ $<$ 2.4 \\
~~ $\geq$6j $\geq$ 3b $\geq$ 3l~\cite{ref-cms-ttbb-diff-13TeV} & ~~1.036 $\pm$ 0.090 $\pm$ 0.012 & 0.899 &  $\pt$ $>$ 25 GeV, $|\eta|$ $<$ 2.4 \\
~~$\geq$ 6j $\geq$ 4b~\cite{ref-cms-ttbb-diff-13TeV} & ~~0.289 $\pm$ 0.036 $\pm$ 0.006 & 0.240 & $\pt$ $>$ 25 GeV, $|\eta|$ $<$ 2.4 \\
~~$\geq$ 7j $\geq$ 4b $\geq$ 3l~\cite{ref-cms-ttbb-diff-13TeV} & ~~0.144 $\pm$ 0.025 $\pm$ 0.005 & 0.129 & $\pt$ $>$ 25 GeV, $|\eta|$ $<$ 2.4 \\
hadronic~\cite{ref-cms-ttbb-hadronic-13TeV}  & 1.6 $\pm$ 0.1$^{+0.5}_{-0.4}$  ~~~~~ & 1.0 $\pm$ 0.2  & $\pt$ $>$ 20 GeV, $|\eta|$ $<$ 2.5  \\
dilepton~\cite{ref-cms-ttcc-13TeV} & ~~ 0.132 $\pm$ 0.010 $\pm$ 0.015 & 0.097 $\pm$ 0.021 & $\pt$ $>$ 20 GeV, $|\eta|$ $<$ 2.4  \\
\midrule
$\ttcc$ \\
dilepton~\cite{ref-cms-ttcc-13TeV} & 0.207 $\pm$ 0.025$\pm$0.027 & 0.187 $\pm$ 0.038 & $\pt$ $>$ 20 GeV, $|\eta|$ $<$ 2.4       \\ 
\bottomrule
\end{tabularx}
\end{threeparttable}
\end{table}
\unskip

\begin{table}[H] 
\caption{
Measured and predicted cross-section ratios in the visible phase space at $\sqrt{s}$ = 13 TeV by CMS. The~predictions of \POWHEG+\PYTHIA8 are shown. 
\label{table:visible_ratio}}
\newcolumntype{C}{>{\centering\arraybackslash}X}
\begin{threeparttable}
\begin{tabularx}{\textwidth}{lCCC}
\toprule
\textbf{Channel}	   & \textbf{Measurements (\textbf{\%})}             & \textbf{Predictions (\textbf{\%})} 	& \textbf{b jet Requirement}\\
\midrule
$\ttbb$/$\ttjj$\\
dilepton~\cite{ref-cms-ttbb-dilepton-13TeV}  & 2.4 $\pm$ 0.3 $\pm$ 0.7   &  1.4 $\pm$ 0.1	    & $\pt$ $>$ 20 GeV, $|\eta|$ $<$ 2.5 \\
dilepton~\cite{ref-cms-ttbb-leptonjets-13TeV}	 & 1.7 $\pm$ 0.1 $\pm$ 0.1   &  1.3 $\pm$ 0.2 	    & $\pt$ $>$ 30 GeV, $|\eta|$ $<$ 2.4 \\
lepton + jets~\cite{ref-cms-ttbb-leptonjets-13TeV} & 2.0 $\pm$ 0.1 $\pm$ 0.1   &  1.7 $\pm$ 0.2	    & $\pt$ $>$ 20 GeV, $|\eta|$ $<$ 2.5 \\
dilepton~\cite{ref-cms-ttcc-13TeV}   & ~~1.93 $\pm$ 0.15 $\pm$ 0.18	&   1.31 $\pm$ 0.12 	& $\pt$ $>$ 20 GeV, $|\eta|$ $<$ 2.4 \\
\midrule
$\ttcc$/$\ttjj$\\
dilepton~\cite{ref-cms-ttcc-13TeV}   & ~~3.01 $\pm$ 0.34 $\pm$ 0.31  &  2.53 $\pm$ 0.18      & $\pt$ $>$ 20 GeV, $|\eta|$ $<$ 2.4 \\
\bottomrule
\end{tabularx}
\end{threeparttable}
\end{table}
\unskip

\begin{table}[H] 
\caption{
Measured and predicted cross-sections in the full phase space at $\sqrt{s}$ = 13 TeV by CMS.
The predictions of \POWHEG+\PYTHIA8 prediction are shown. 
\label{table:full_cross}}
\newcolumntype{C}{>{\centering\arraybackslash}X}
\begin{threeparttable}
\begin{tabularx}{\textwidth}{lCCC}
\toprule
\textbf{Channel}	   & \textbf{Measurements (\textbf{pb}) }            & \textbf{Predictions (\textbf{pb})} 	& \textbf{b jet Requirement}\\
\midrule
$\ttbb$\\
dilepton~\cite{ref-cms-ttbb-dilepton-13TeV}   & ~~~~4.0 $\pm$ 0.6 $\pm$ 1.3   &  3.2 $\pm$ 0.4	    & $\pt$ $>$ 20 GeV, $|\eta|$ $<$ 2.5 \\
dilepton~\cite{ref-cms-ttbb-leptonjets-13TeV}	 & ~~~~2.9 $\pm$ 0.1 $\pm$ 0.5   &  2.3 $\pm$ 0.4 	    & $\pt$ $>$ 30 GeV, $|\eta|$ $<$ 2.4 \\
lepton + jets~\cite{ref-cms-ttbb-leptonjets-13TeV} & ~~~~4.7 $\pm$ 0.2 $\pm$ 0.6   &  3.9 $\pm$ 0.4	    & $\pt$ $>$ 20 GeV, $|\eta|$ $<$ 2.5 \\
hadronic~\cite{ref-cms-ttbb-hadronic-13TeV}    & 5.5 $\pm$ 0.3$^{+1.6}_{-1.3}$   & 3.5 $\pm$ 0.6  	& $\pt$ $>$ 20 GeV, $|\eta|$ $<$ 2.5 \\
dilepton~\cite{ref-cms-ttcc-13TeV}   & ~~~~~~4.54 $\pm$ 0.34 $\pm$ 0.56	&   3.34 $\pm$ 0.72 	& $\pt$ $>$ 20 GeV, $|\eta|$ $<$ 2.4 \\
\midrule
$\ttcc$\\
dilepton~\cite{ref-cms-ttcc-13TeV}   & ~~10.1 $\pm$ 1.2 $\pm$ 1.4  &  9.1 $\pm$ 1.8      & $\pt$ $>$ 20 GeV, $|\eta|$ $<$ 2.4 \\
\bottomrule
\end{tabularx}
\end{threeparttable}
\end{table}
\unskip

\begin{table}[H] 
\caption{
Measured and predicted cross-section ratios in the full phase space at $\sqrt{s}$ = 13 TeV by CMS.
The predictions of \POWHEG+\PYTHIA8 prediction are shown.
\label{table:full_ratio}}
\newcolumntype{C}{>{\centering\arraybackslash}X}
\begin{threeparttable}
\begin{tabularx}{\textwidth}{lCCC}
\toprule
\textbf{Channel}	   & \textbf{Measurements (\textbf{\%})}             & \textbf{Predictions (\textbf{\%})} 	& \textbf{b jet Requirement}\\
\midrule
$\ttbb$/$\ttjj$\\
dilepton~\cite{ref-cms-ttbb-dilepton-13TeV}   & 2.2 $\pm$ 0.3 $\pm$ 0.6   &  1.2 $\pm$ 0.1	    & $\pt$ $>$ 20 GeV, $|\eta|$ $<$ 2.5 \\
dilepton~\cite{ref-cms-ttbb-leptonjets-13TeV}	 & 1.8 $\pm$ 0.1 $\pm$ 0.2   &  1.4 $\pm$ 0.3 	    & $\pt$ $>$ 30 GeV, $|\eta|$ $<$ 2.4 \\
lepton + jets~\cite{ref-cms-ttbb-leptonjets-13TeV} & 1.6 $\pm$ 0.1 $\pm$ 0.1   &  1.3 $\pm$ 0.2	    & $\pt$ $>$ 20 GeV, $|\eta|$ $<$ 2.5 \\
dilepton~\cite{ref-cms-ttcc-13TeV}   & ~~1.51 $\pm$ 0.11 $\pm$ 0.16	&   1.03 $\pm$ 0.08 	& $\pt$ $>$ 20 GeV, $|\eta|$ $<$ 2.4 \\
\midrule
$\ttcc$/$\ttjj$\\
dilepton~\cite{ref-cms-ttcc-13TeV}   & ~~3.36 $\pm$ 0.38 $\pm$ 0.34  &  2.81 $\pm$ 0.20      & $\pt$ $>$ 20 GeV, $|\eta|$ $<$ 2.4 \\
\bottomrule
\end{tabularx}
\end{threeparttable}
\end{table}
\vspace{-14pt}

\unskip  

\subsection{Differential Cross-Section~Measurements}\label{exp-diff}

In addition to the inclusive cross-section measurements, the~differential measurements of the $\ttbar$ + HF production cross-sections 
can also provide information on the perturbative QCD (pQCD) and enable the searches for potential new physics.
The $\ttbb$ differential cross-section measurements have been performed at $\sqrt{s}$ = 7, 8 and 13 TeV with the ATLAS experiment and at $\sqrt{s}$ = 8 and 13 TeV with the CMS~experiment.

To measure the differential cross-sections, the~measured distributions at the detector level need to be unfolded to the generator level where the detector effect is removed so that the resulting cross-section can be compared with theory predictions and results from other experiments.
At the generator level, it is not trivial to define the additional b jets in the $\ttbb$ process as we have b jets from the top quark decay. Moreover, the~b jet could also emerge from the W boson decay. The~additional b jets are expected to come from the gluon decay and can also come from the decay of the H boson or another~boson. 

In the b jet identification, there is a clear difference between the two experiments in ATLAS and CMS. 
In the ATLAS experiment, at~the particle level, there is no attempt to identify the origin of the b jets relying on the simulation information. 
At this particle level, the~two b jets with the highest $\pt$ or the smallest $\Delta R$ are selected for the differential cross-section measurement. 
The highest $\pt$ jets are considered as the b jets from the top quark while the b jets with the smallest $\Delta R$ are considered as the additional b jets not from the top quark decay to make use of the fact that the b jets from a gluon splitting tend to be collinear. 
While in CMS, the~origin of the b jets is explicitly identified using the simulation information. For~example, the~b hadron is traced back through its ancestors in the simulation chain. In~this way, only if the b jet is not from a top quark, the~b jet is identified as one of the two additional b~jets. 

For the ATLAS measurements, the~unfolded results are presented as normalized differential cross-sections in visible phase space as a function of the b jet multiplicity, global event properties and various kinematic variables. The~measurements are conducted in the $e\mu$ channel with at least three reconstructed b jets and in the lepton + jets channel with at least four b jets.
The sample with at least four b jets in the lepton + jets channel has high signal purity resulting in a measurement with smaller dependence on the simulation. The~$e\mu$ channel benefits from an order of magnitude of a larger sample size containing at least three b~jets. 

\begin{figure}[H]
\includegraphics[width=11.5 cm]{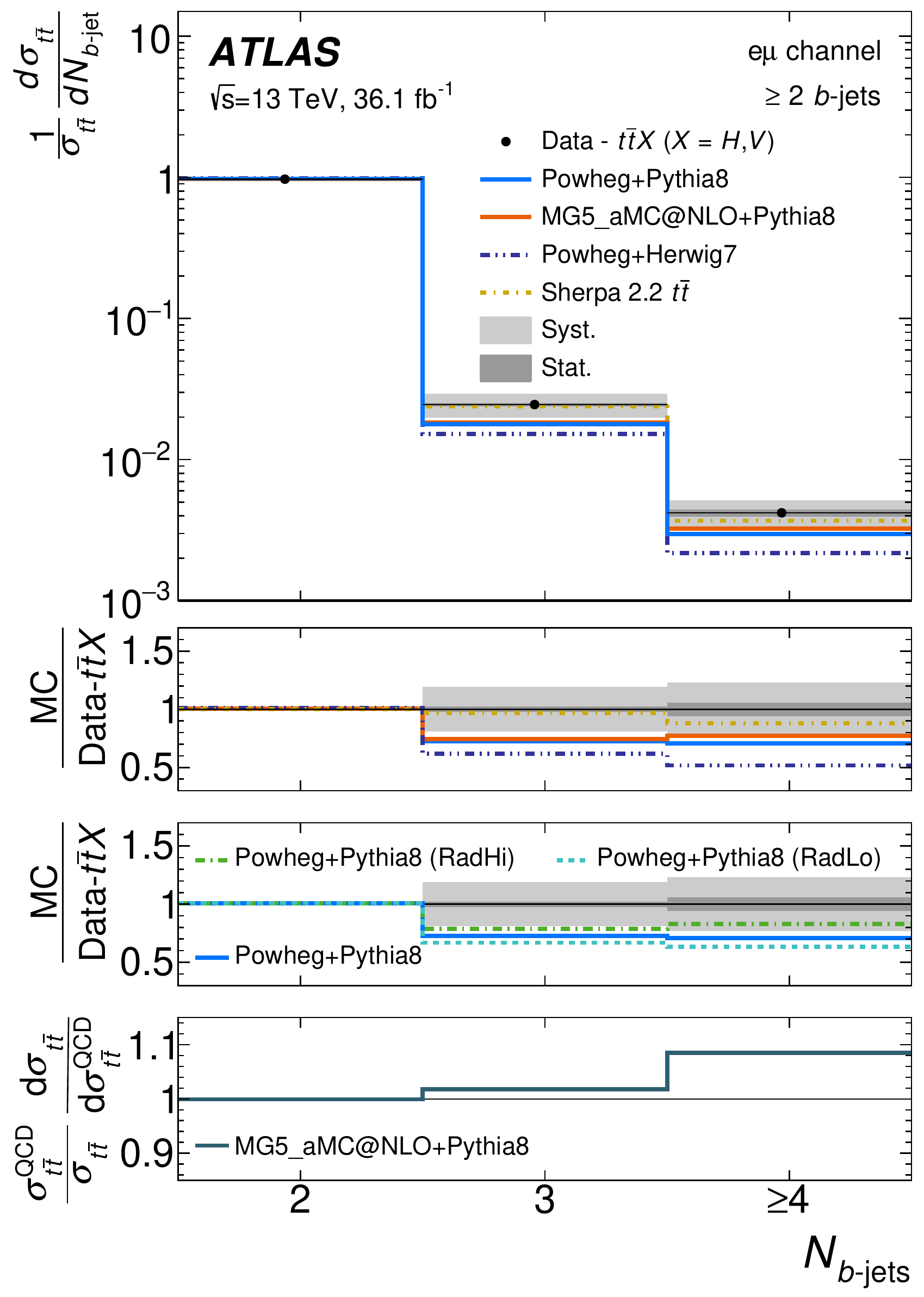}
\caption{{The} 
 relative differential cross-section as a function of the b jet multiplicity in events with at least two b jets in the $e\mu$ channel compared with various generators. The~$\ttH$ and $\ttV$ contributions are subtracted from data. 
Uncertainty bands represent the statistical and total systematic uncertainties~\cite{ref-atlas-13TeV}.\label{atlas_13TeV_fig_08}}
\end{figure}

Once the reconstructed level distributions of $\ttbar$ + HF events are extracted, then
the measured distributions are unfolded to the particle level. 
The detector resolution effect and inefficiency are corrected by inverting the migration matrix which is optimized for a diagonal matrix.  
An iterative Bayesian unfolding technique~\cite{bayesian-unfold} implemented in the \ROOUNFOLD
software package~\cite{roounfold} is used in this process.
Detector efficiencies and acceptance are then corrected using a bin-by-bin method.  
Figure~\ref{atlas_13TeV_fig_08} shows the normalized cross-section as a function of the b jet multiplicity compared with predictions from various generator set-ups. The~first three panels show the ratios of various predictions to data. The~last panel shows the ratio of predictions of normalized differential cross-sections from \MADGRAPH 5\_aMC@NLO+\PYTHIA 8, including or not the contributions from the $\ttH$ and $\ttV$ processes. All predictions relying on the parton shower generation of jets for high multiplicities are lower compared to the measurements.
This suggests that the b jet production by the parton shower is not optimal in these processes. The~comparison of the predictions from various generators with the measurements are made after subtracting the simulation-estimated contributions of $\ttV$ and $\ttH$ production from the data. The~impact of including these processes in the prediction increases with b jet multiplicity, resulting in a change of about 10\% relative to the QCD $\ttbar$ prediction alone in the inclusive four b jet bin. The~measurement in the $e\mu$ channel with at least three b jets tends to be more precise than in the lepton + jets channel with at least four b~jets. 

\begin{figure}[H]
\includegraphics[width=6.85 cm]{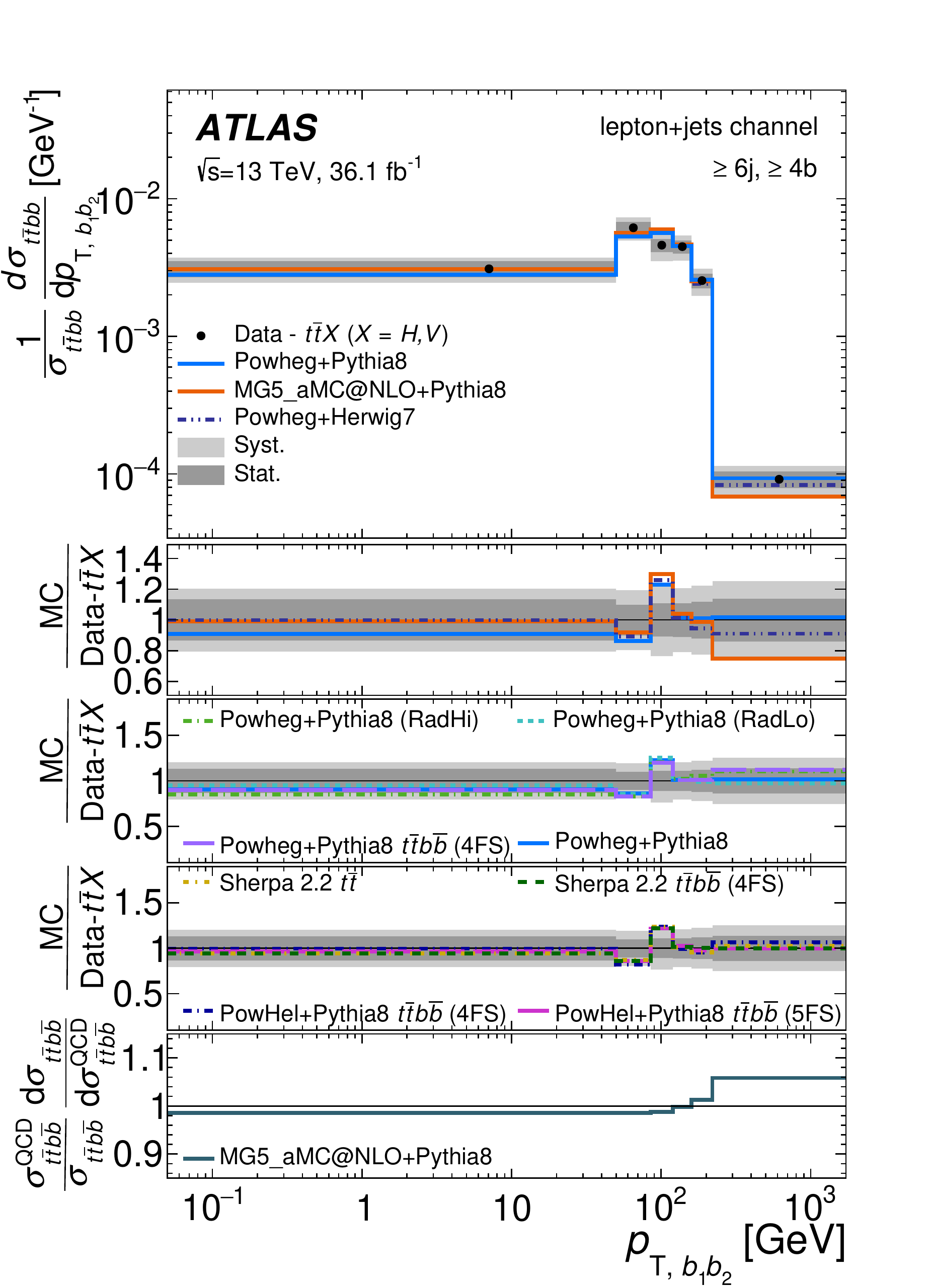}
\includegraphics[width=6.85 cm]{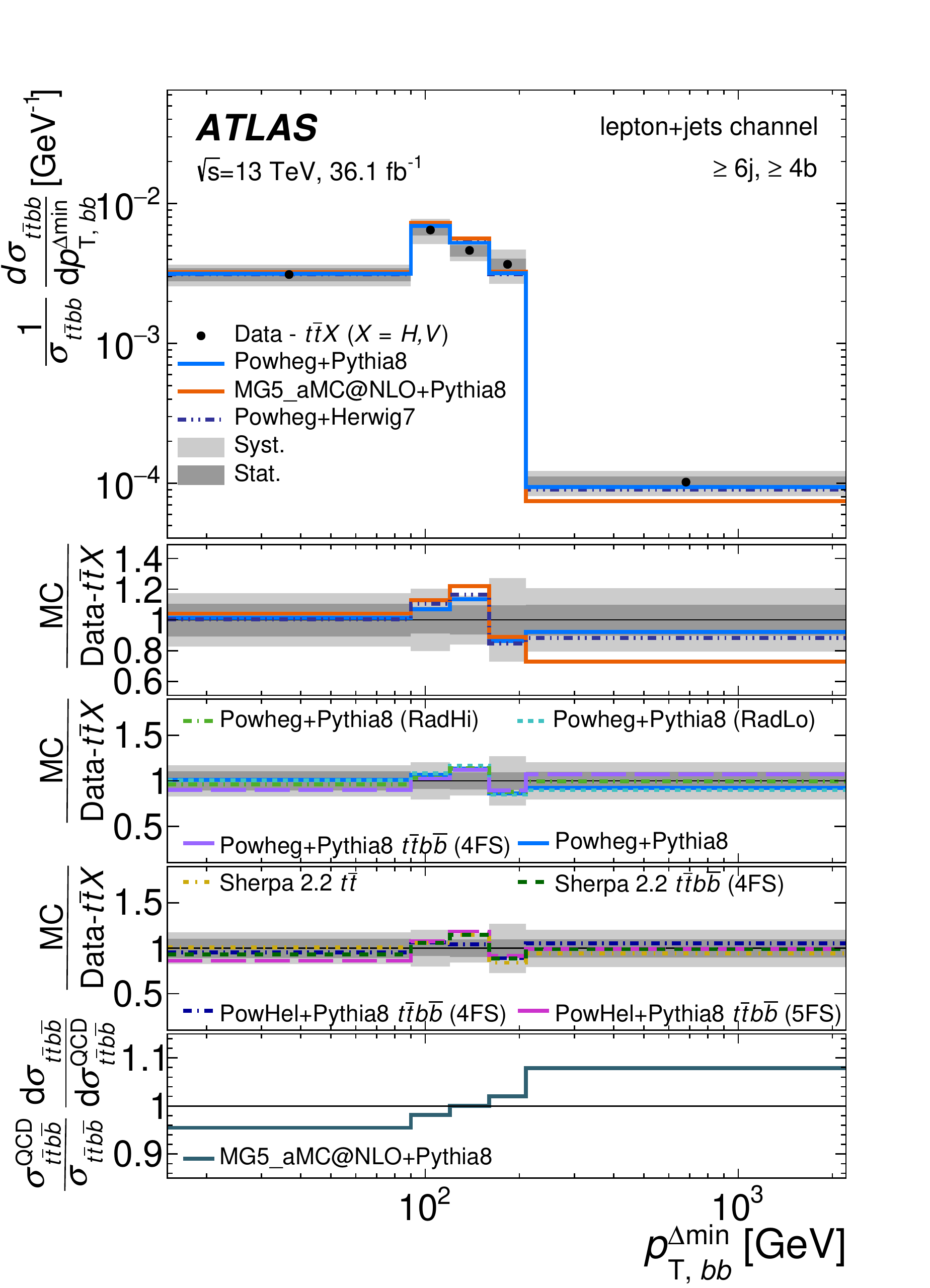}
\caption{{Relative} 
 differential cross-sections as a function of $\pt$ of the two highest-$\pt$ b jets (\textbf{left}) and the two closest b jets in $\Delta R$ (\textbf{right}) in the events with at least four b jets in the lepton + jets channel compared with various generators from the ATLAS measurements. The~contributions from $\ttH$ and $\ttV$ are subtracted from data~\cite{ref-atlas-13TeV}.\label{atlas_13TeV_pt}
}
\end{figure}

\begin{figure}[H]
\includegraphics[width=6.7 cm]{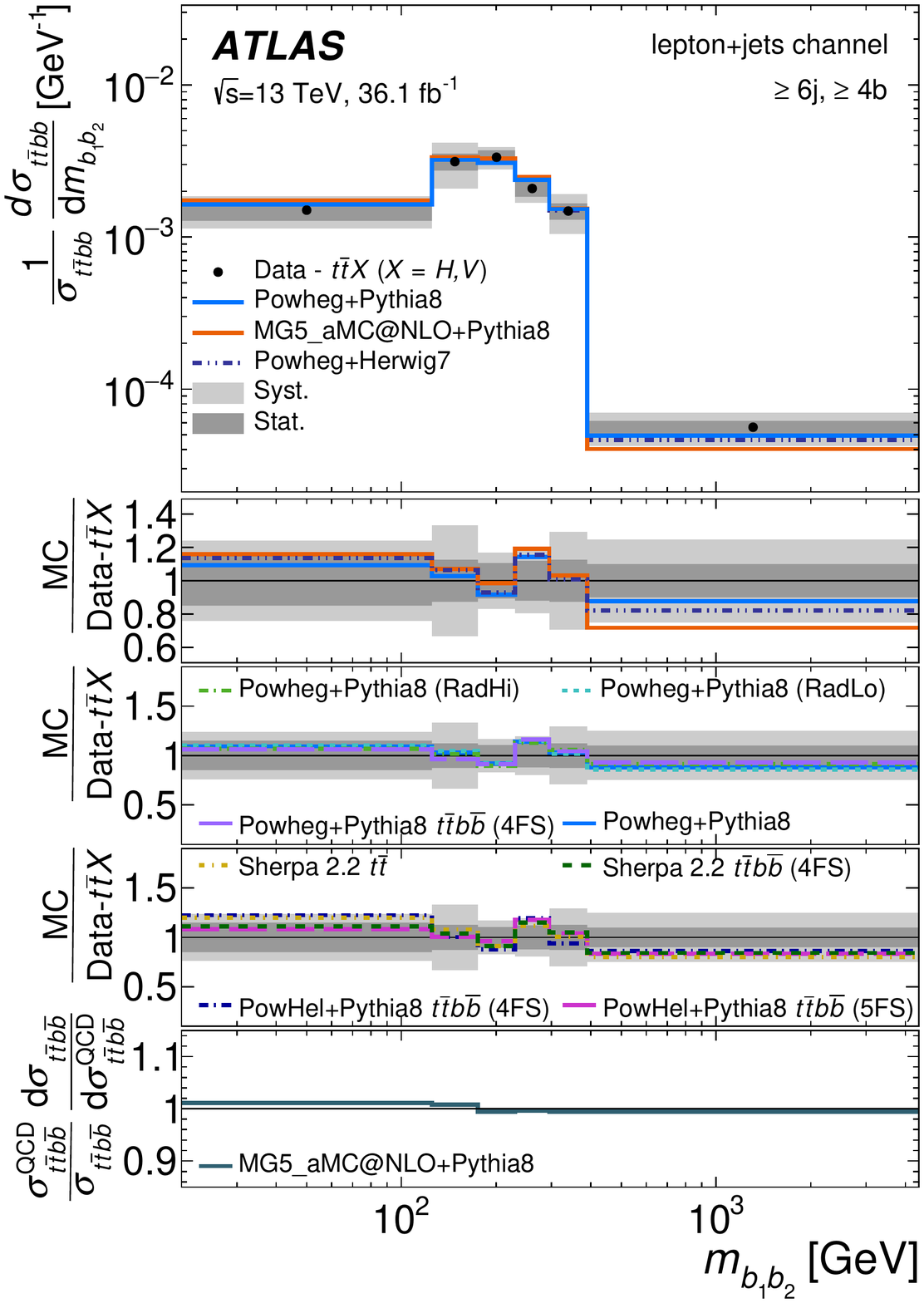}
\includegraphics[width=6.7 cm]{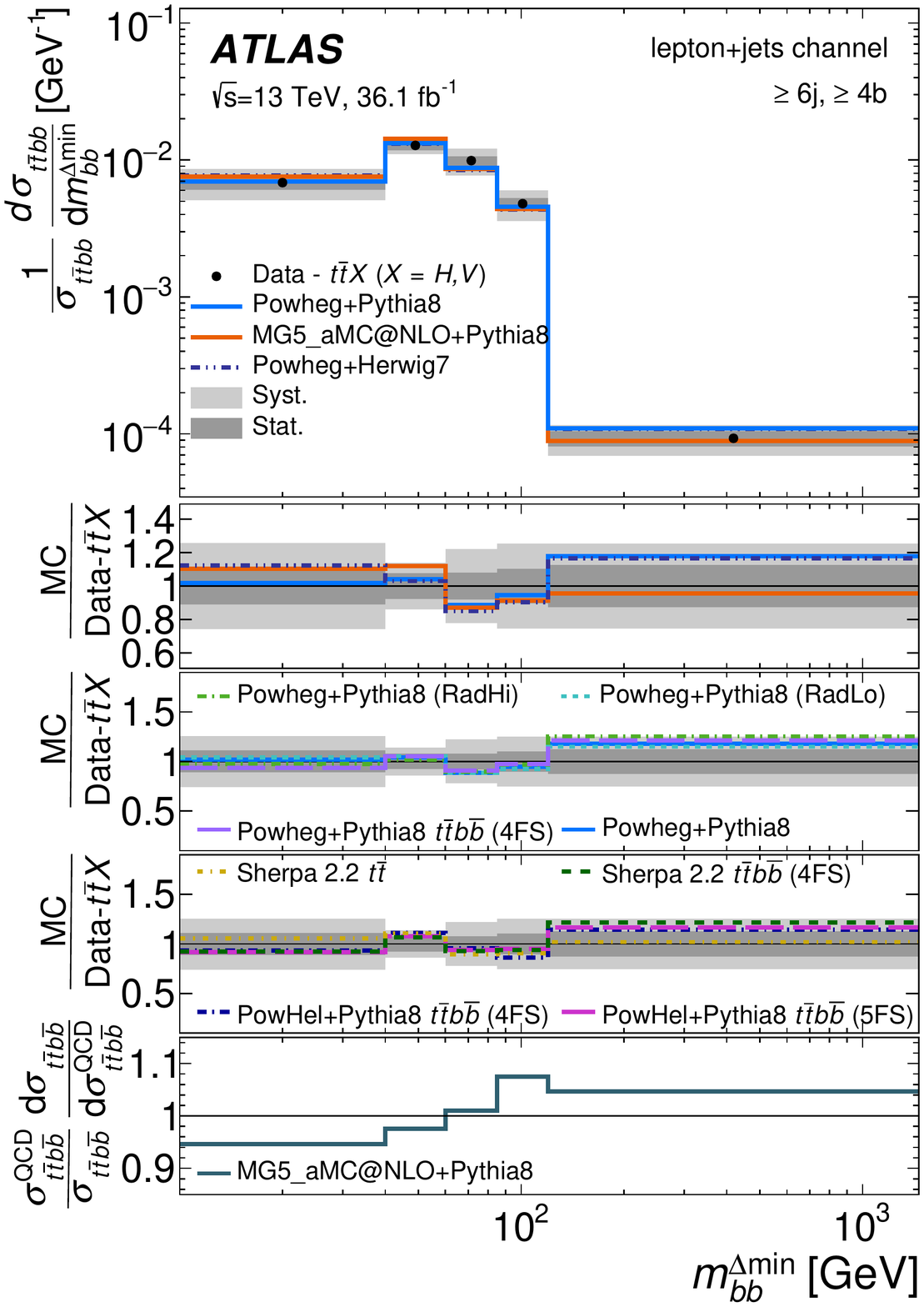}
\caption{{Relative} 
 differential cross-sections as a function of $m_{b_1 b_2}$ of the two highest-$\pt$ b jets (\textbf{left}) and the two closest b jets in $\Delta R$ (\textbf{right}) in the events with at least four b jets in the lepton + jets channel compared with various generators from the ATLAS measurements. The~contributions from $\ttH$ and $\ttV$ are subtracted from data~\cite{ref-atlas-13TeV}.\label{atlas_13TeV_mass}
}
\end{figure}

\begin{figure}[H]
\includegraphics[width=6.85 cm]{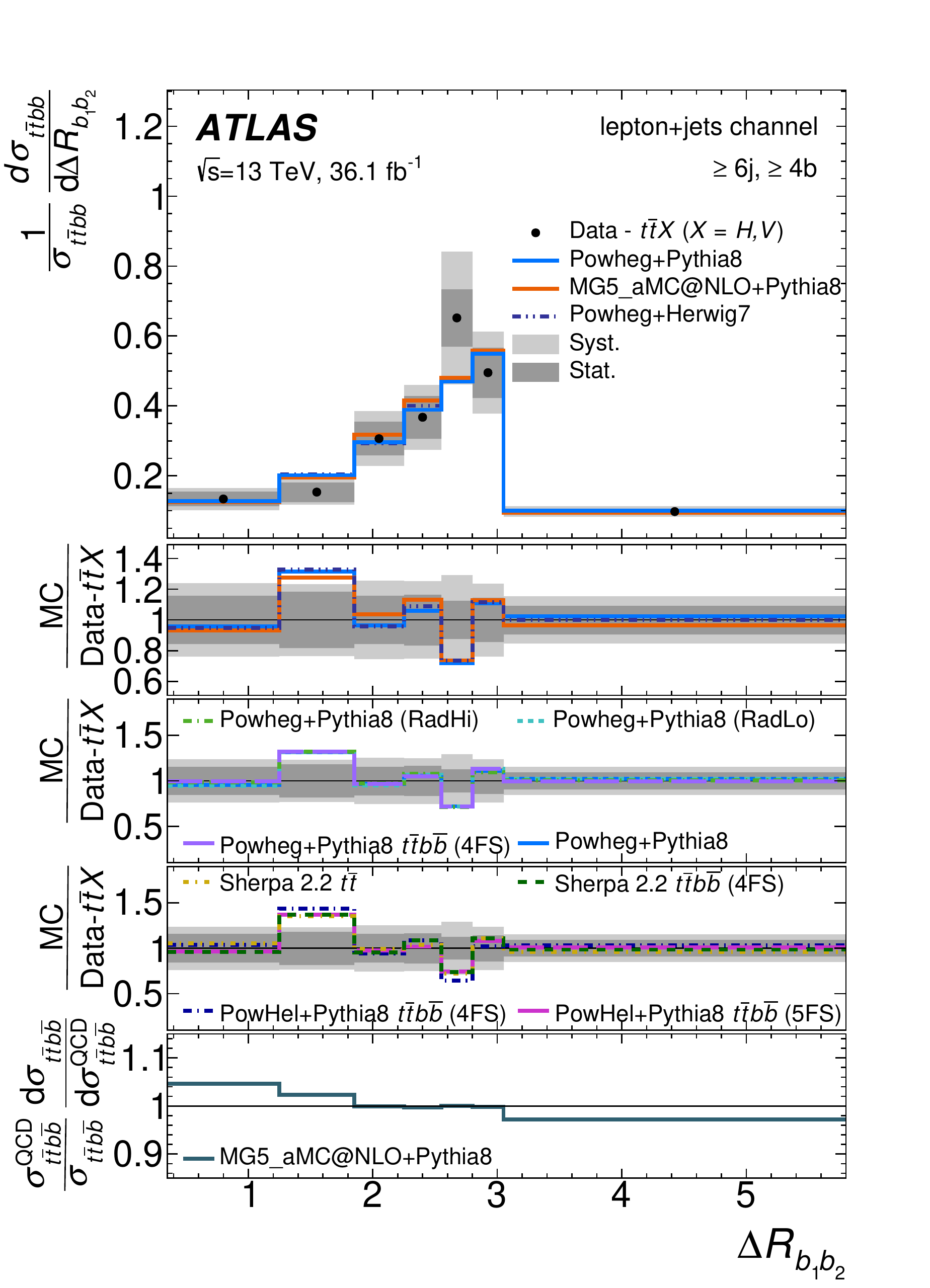}
\includegraphics[width=6.85 cm]{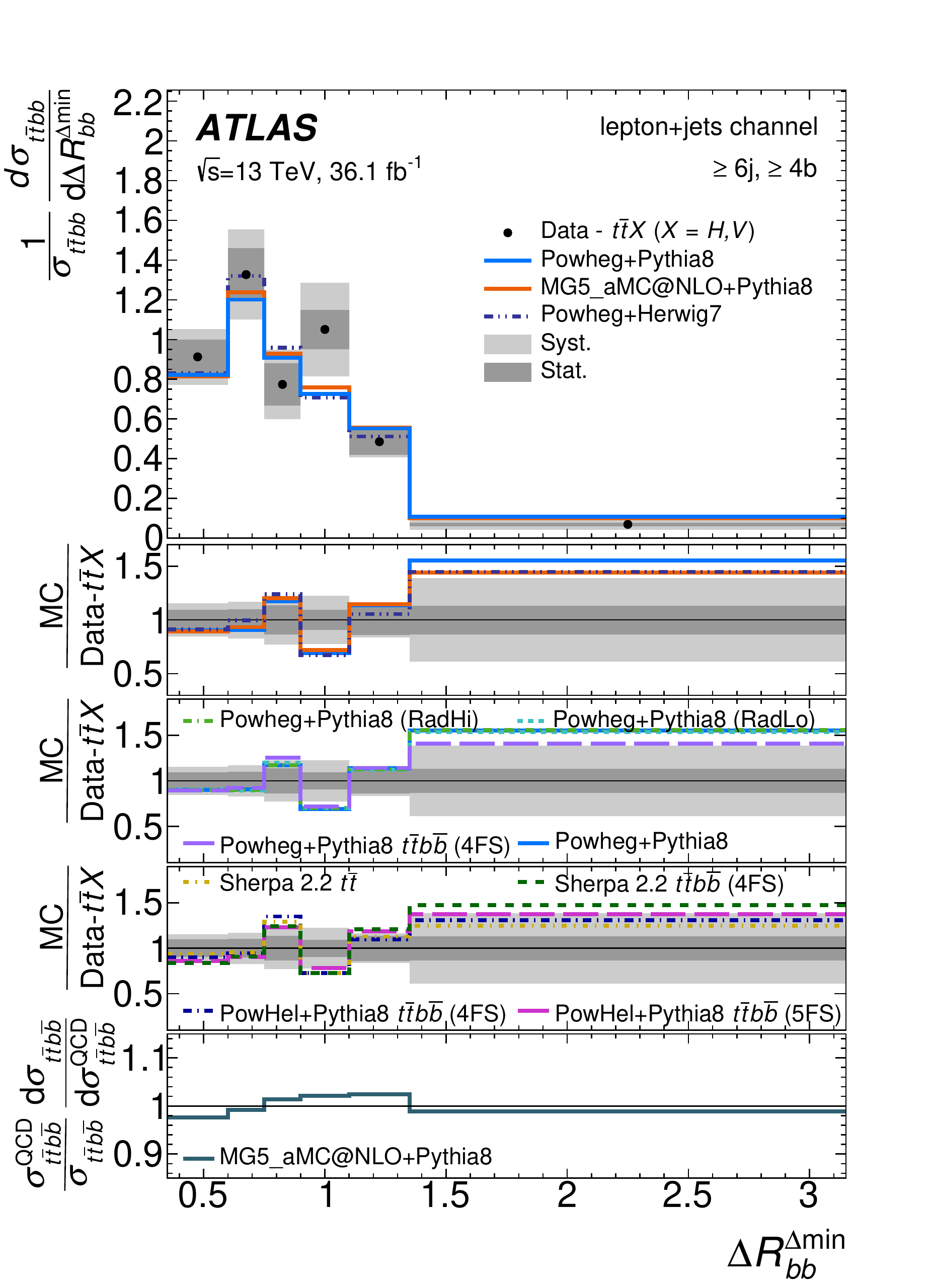}
\caption{{Relative} 
 differential cross-sections as a function of $\Delta R_{b_1,b_2}$ of the two highest$-$$\pt$ b jets~(\textbf{left}) and the two closest b jets in $\Delta R$ (\textbf{right}) in the events with at least four b jets in the lepton + jets channel compared with various generators from the ATLAS measurements. The~contributions from $\ttH$ and $\ttV$ are subtracted from data~\cite{ref-atlas-13TeV}.\label{atlas_13TeV_dr}
}
\end{figure}

It is also of importance to verify the distributions of the $\pt$, the~mass and the angular distance $\Delta R$ of the two b jets where the $b_1 b_2$ system is built from the two highest-$\pt$ b jets and the two closest b jets in $\Delta R$. The~measured distributions of those three variables in the lepton + jets channel are shown in Figures~\ref{atlas_13TeV_pt}--\ref{atlas_13TeV_dr}.
The differential cross-section as a function of the $\pt$ of the $b_1 b_2$ system is measured with a precision of 10--15\% over the full range in the $e\mu$ channel and with an uncertainty of 20--25\% in the lepton + jets channel. 
In general, the~differential distributions are well described by the different theoretical predictions, which vary significantly less compared to the size of the experimental uncertainty. 
All other distributions such as $H_T$ or $\pt$ of additional b jets are available in Ref.~\cite{ref-atlas-13TeV}.

In CMS, the~differential cross-sections are measured in the visible phase space as a function of various kinematic properties such as the $\pt$ and $\eta$ of the leading and subleading additional b jets, the~angular distance $\Delta R$ between them and 
the invariant mass $m_{\bbbar}$ of the two additional b jets. In~particular, the~differential cross-sections as a function of the $m_{\bbbar}$ and $\Delta R$ are of interest as the two additional b jets from a gluon tend to be produced collinearly and those from the H boson have the resonance peak at 125~GeV.

\vspace{-3pt}
\unskip  

\vspace{-12pt}

At the reconstruction level, it is very challenging to identify two additional b jets because there are four b jets from top quarks and a gluon splitting.  
To select the additional b jets, the~multivariate approach of a BDT was used to maximize the correct assignment of additional b jets.
The input variables to the BDT combine information from the two final-state leptons, the~jets and~$E_T^{miss}$.
A total of twelve variables, e.g.,~the sum and difference of the invariant mass of the $bl^+$ and $\bar{b}l^-$ system, $m^{bl^+}\pm m^{\bar{b}l^-}$; the absolute difference in the azimuthal angle between them, $|\Delta \phi^{bl^+,\bar{b}l^-}|$; the $\pt$ of the $bl^+$ and $\bar{b}l^-$ system, $\pt^{bl^+}$ and $\pt^{\bar{b}l^-}$ and the difference between the invariant mass of the two b jets and two leptons and the invariant mass of the $\bbbar$ pair, $m^{b\bar{b} l^+l^-}-m^{b\bar{b}}$, are used as input variables. The variables insensitive to the additional radiation are selected to avoid any dependence on the kinematics of the additional jets.
The jets from the $\ttbar$ system are identified as the pair with the highest BDT discriminant.
From the remaining jets, those b-tagged jets with the highest $\pt$ are selected as being the leading additional ones. 
With this method, the~correct assignment rate for the additional b jets in $\ttbb$ events is around 40\%. 

\begin{figure}[H]
\includegraphics[width=6.85 cm]{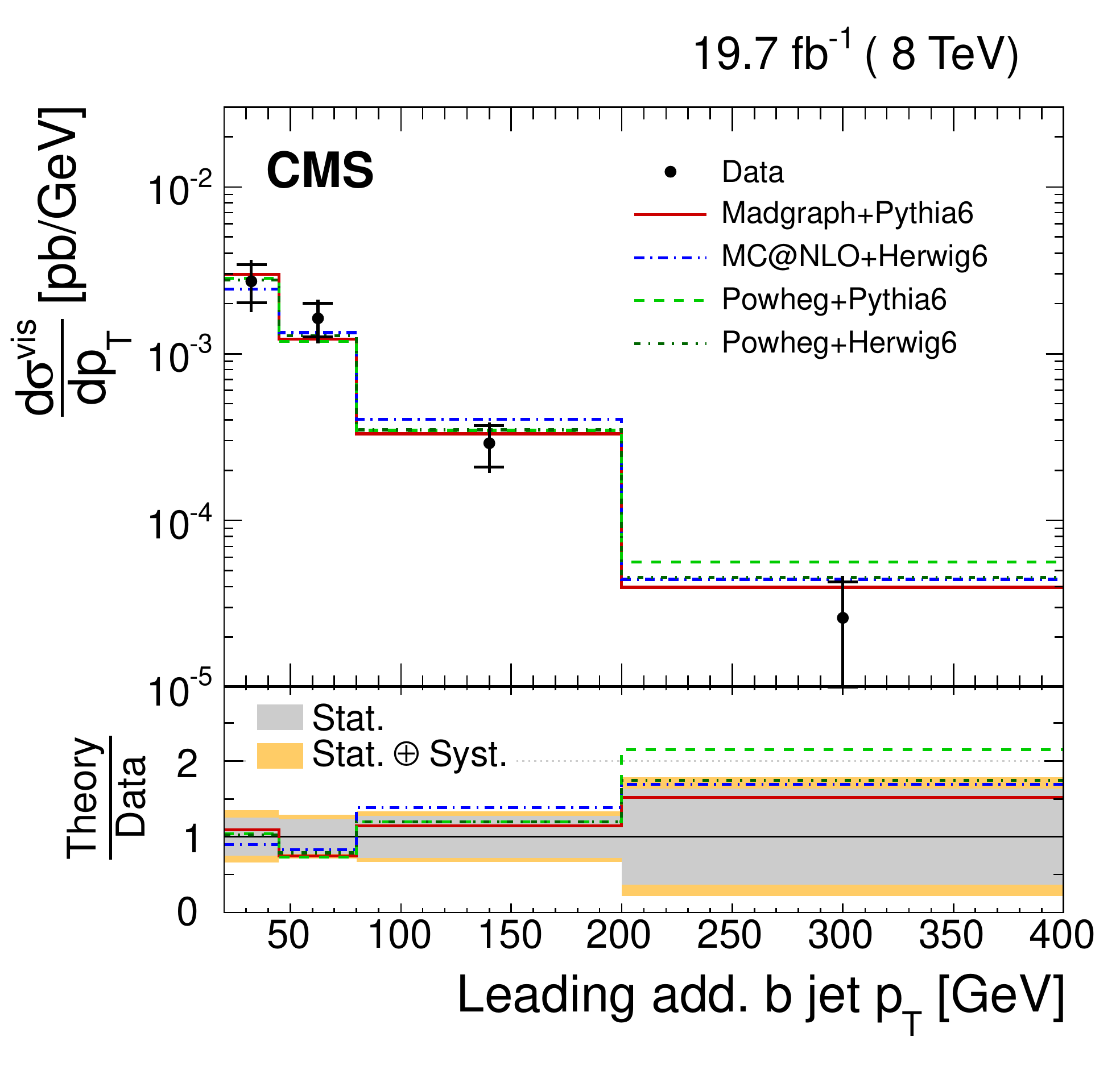}
\includegraphics[width=6.85 cm]{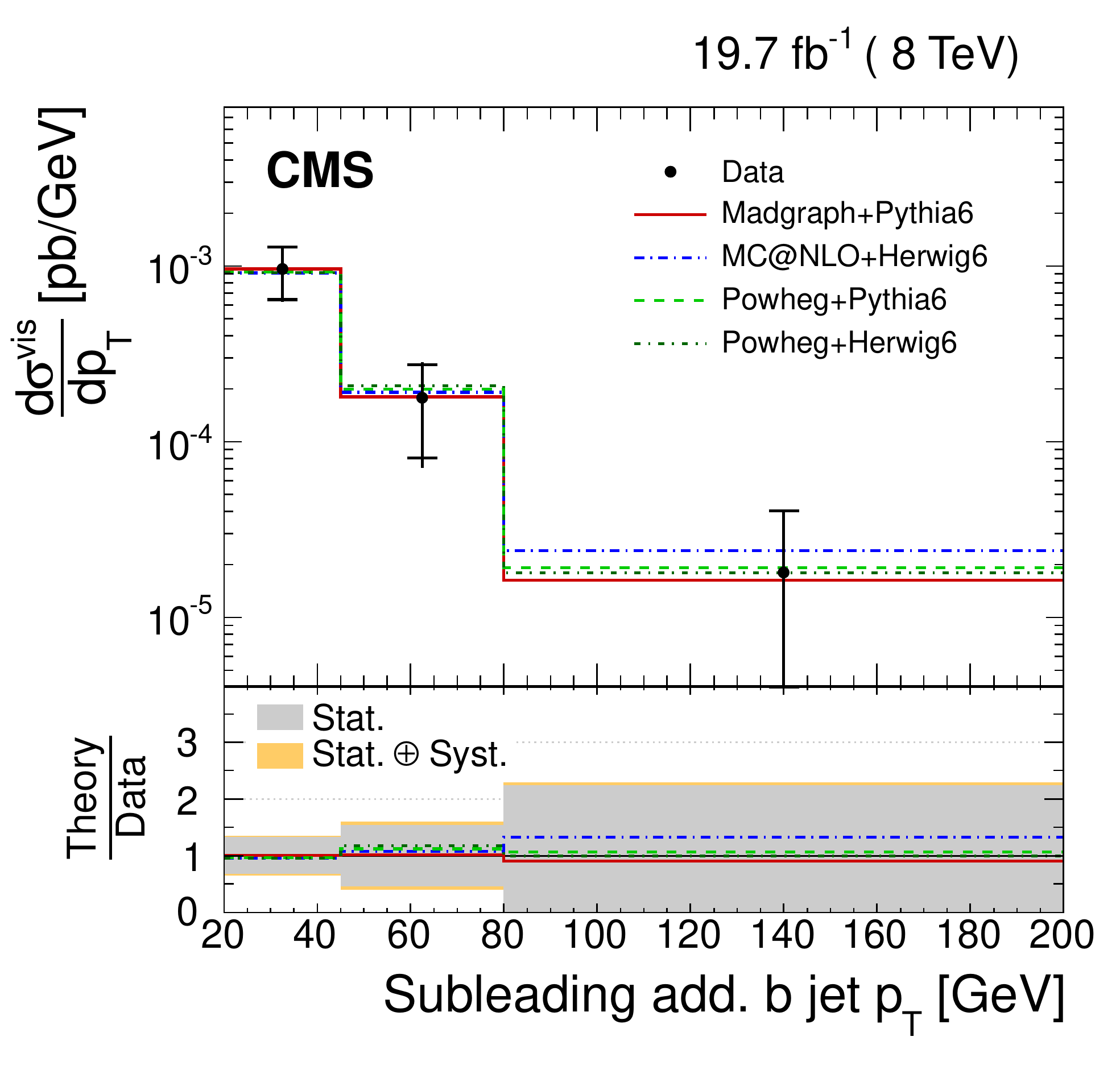}\\
\includegraphics[width=6.85 cm]{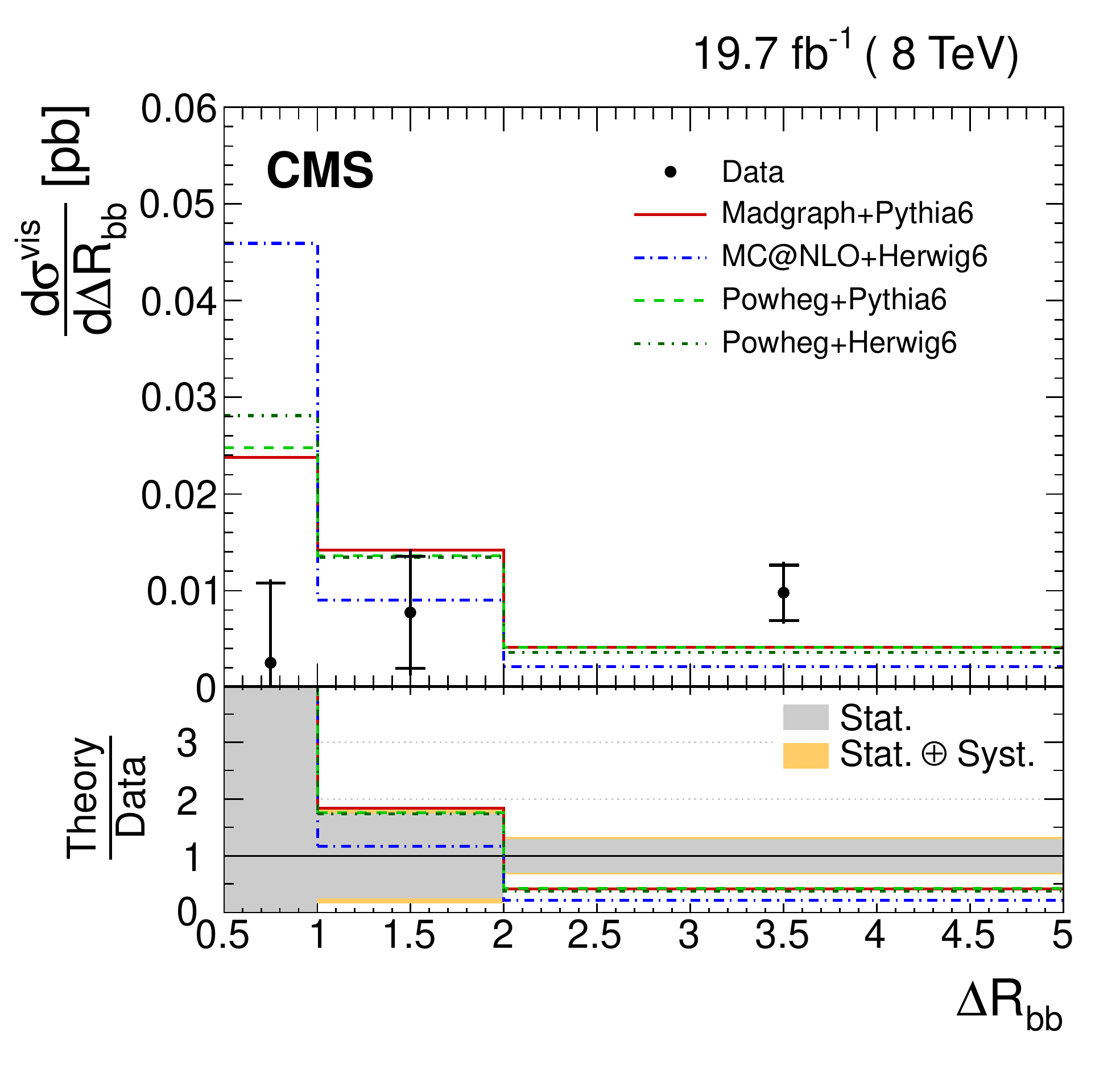}
\includegraphics[width=6.85 cm]{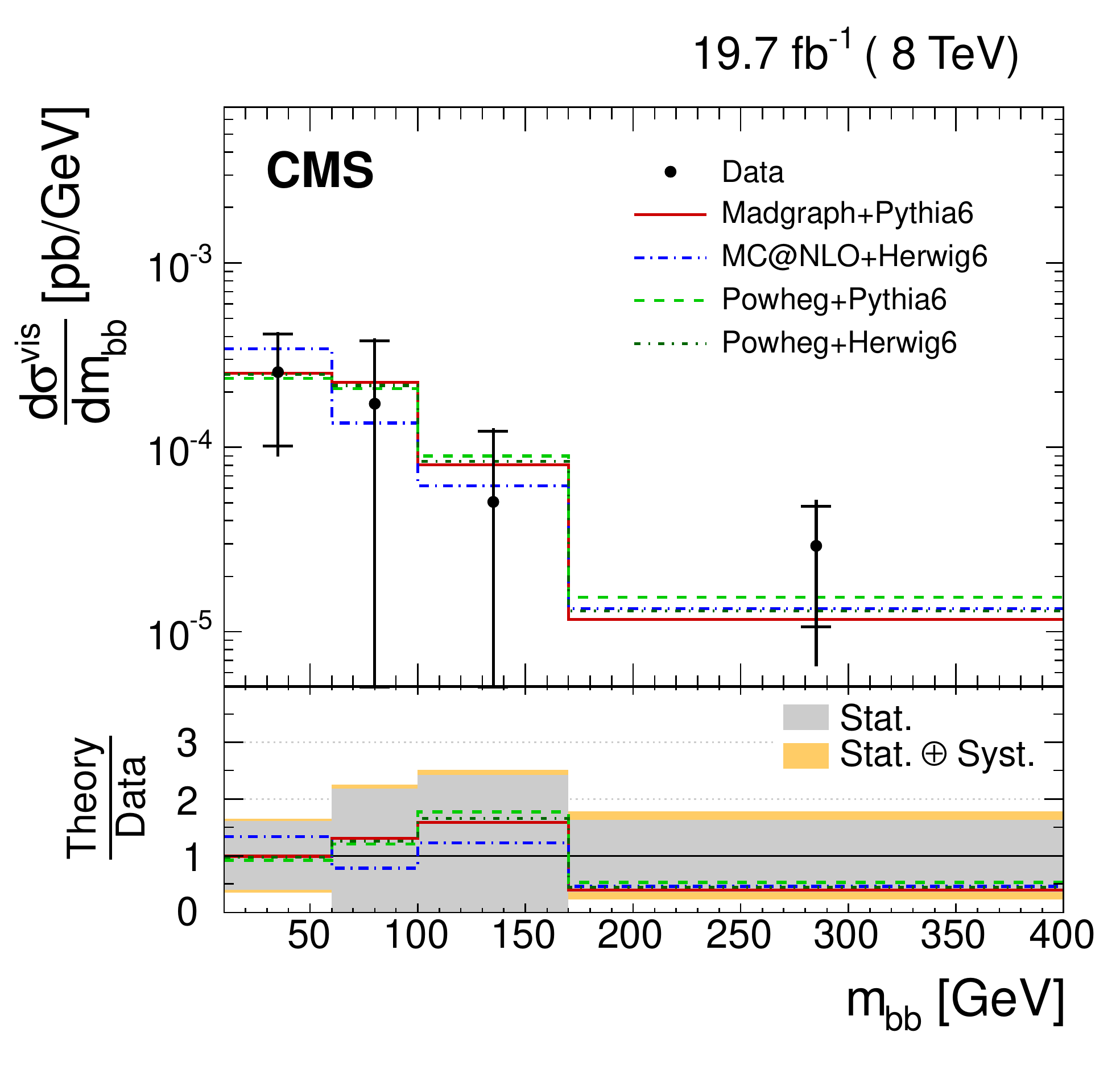}\vspace{-6pt}
\caption{{Absolute} 
 differential $\ttbar$ cross-sections as a function of the leading and subleading additional b jet $\pt$ (\textbf{upper}) and angular separation and invariant mass of two additional b jets (\textbf{lower}) in the visible phase space at the CMS. Inner vertical bars indicate the statistical uncertainties~\cite{ref-cms-ttbb-diff-8TeV}.\label{CMS-TOP-12-041_Figure}}
\end{figure}

A template fit to the b-tagged jet multiplicity distribution is performed to improve the data and simulation comparison. 
For the differential cross-section measurements, effects from detector efficiency and resolution are corrected by using the regularized inversion of the response matrix which is calculated from simulated $\ttbar$ events. 
The measured differential cross-sections as a function of the leading and subleading additional b jet $\pt$, the~$\Delta R$ and invariant mass of two additional b jets are shown in Figure~\ref{CMS-TOP-12-041_Figure} for CMS. 
Measured cross-sections are compared with various theoretical predictions. 
The shape of the $\pt$ distributions are well described by prediction. 
However, the~measured values by CMS have larger uncertainties due to the use of a smaller data sample with respect to~ATLAS. 

In~CMS, the~differential cross-sections are measured with a full Run 2 data in the lepton~+~jets channel. In~this analysis, two approaches are used to identify the additional b jets from the gluon splitting, while two b jets with the smallest angular separation are selected to reduce the systematic uncertainty on theory dependence, 
a~multivariate algorithm based on a deep neural network (DNN) is also used to identify additional b jets not from top quarks by using the MC~information. 

To find the correct pair of b jets not from top quarks, only four b jets in the highest $\pt$ order are used as candidate jets, which results in the six possible candidate jet combinations.  
The DNN makes use of two sets of input variables, targeting jet-specific input information and global event information separately.
For jet-specific input information, 
the input variables consist of the $\pt$, $\eta$, a~flag
indicating whether it passes the tight b tagging working point, the~angular separation ($\Delta R$)
with the charged lepton and~the invariant mass with the charged lepton. These input variables are connected via five convolutional network layers (CNN)~\cite{cnn} followed by a long short-term memory (LSTM) cell~\cite{lstm}.
For the global event information, the~input variables consist of the scalar $\pt$ sum of the four candidate b jets, the~$\pt$, $\eta$, $\phi$ of the charged lepton, the~$\Delta \phi$, $\Delta \eta$ and invariant mass of the dijet combinations, the~$\Delta$ R of the dijet combinations and the charged lepton as well as~the jet and b-tagged jet multiplicities. These input variables are connected to three dense network layers with 50~nodes each. 
Both of these sequences are concatenated at the end into one dense layer with 10 nodes, which is connected
to an output layer consisting of six nodes, each representing one of the six possible candidate jet combinations. 
The pair of b-tagged jets with the highest DNN output value per event is
chosen as the correct assignment of the additional b jet pair and used further for the differential cross-section~measurement. 

\begin{figure}[H]
\includegraphics[width=6.85 cm]{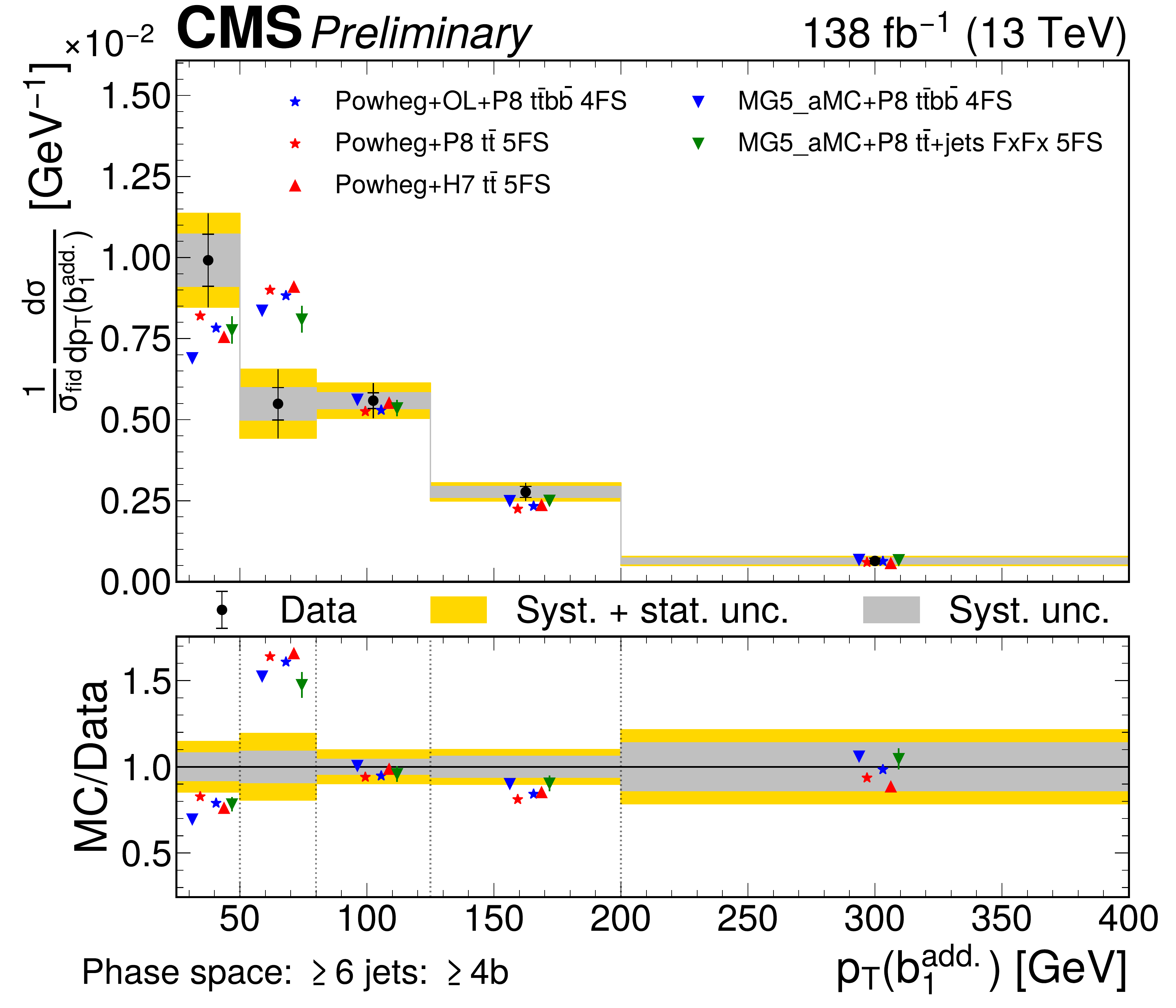}
\includegraphics[width=6.85 cm]{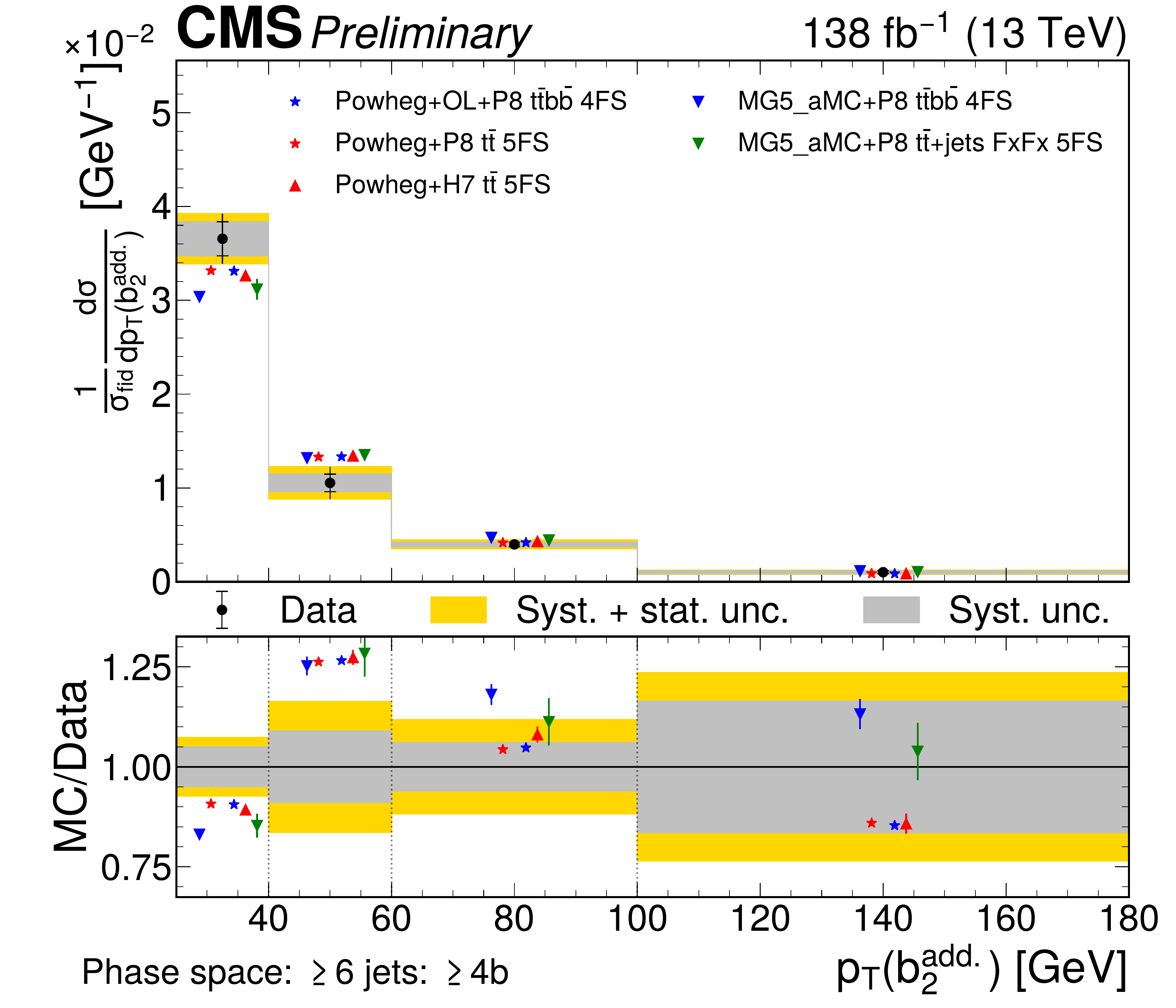}\\
\includegraphics[width=6.85 cm]{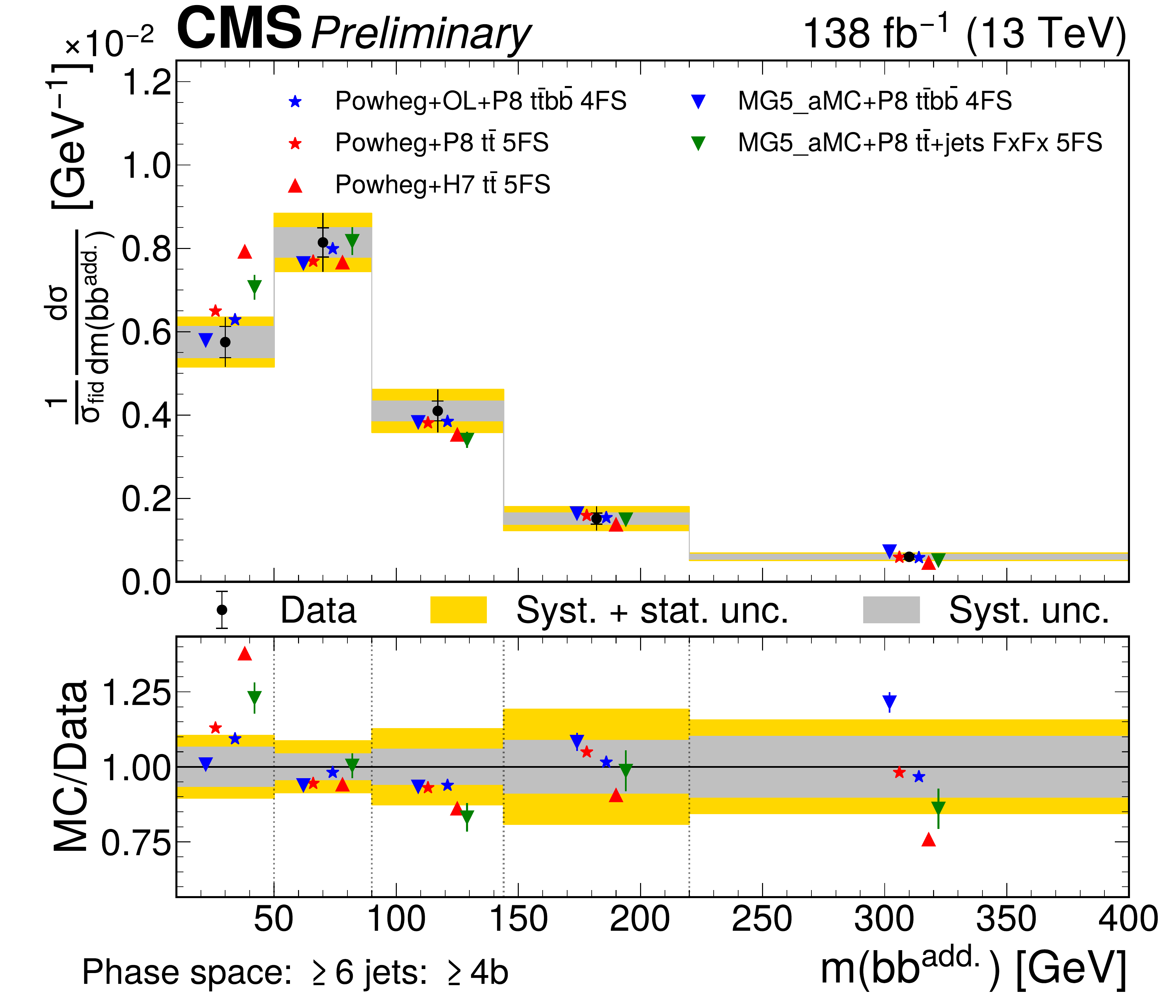}
\includegraphics[width=6.85 cm]{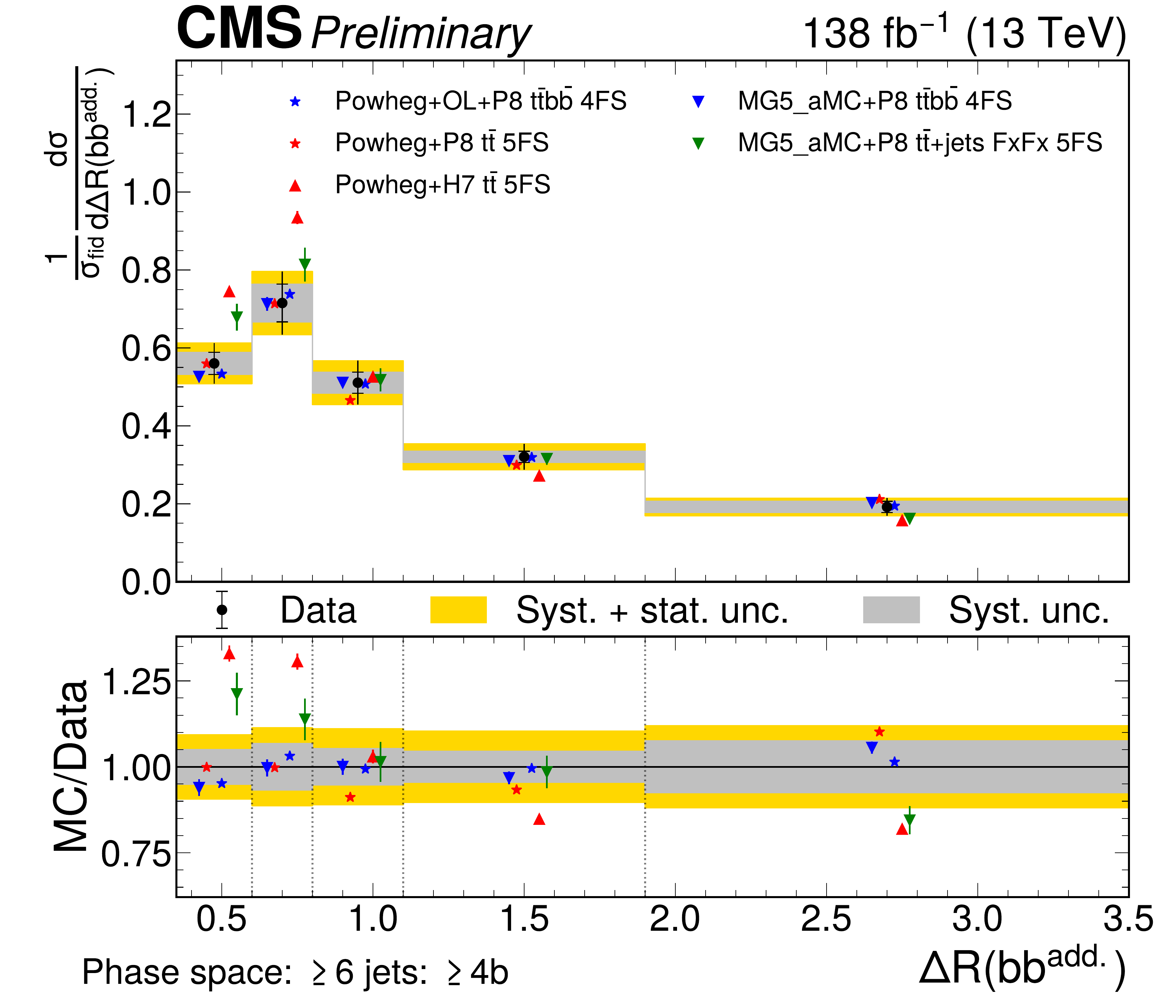}
\caption{Normalized differential $\ttbar$ cross-sections as a function of the leading and subleading additional b jet $\pt$ (\textbf{upper}) and angular separation and invariant mass of two additional b jets (\textbf{lower}) in the visible phase space at the CMS. Inner vertical bars indicate the statistical uncertainties~\cite{ref-cms-ttbb-diff-13TeV}.\label{CMS-TOP-22-009_Figure}}
\end{figure}

The correct assignment of additional b jets in the DNN is about 49\%, which represents a significant increase compared to choosing the two b jets closest in $\Delta R$, which only yields about 41\%. 
The measured differential cross-sections as a function of the leading and subleading additional b jet $\pt$, the~$\Delta R$ and invariant mass of two additional b jets selected in the DNN are shown in Figure~\ref{CMS-TOP-22-009_Figure}. The~distributions are not well described by \mbox{\POWHEG + \HERWIG 7} (referred to as \POWHEG + \textsc{h}7 in Figure~\ref{CMS-TOP-22-009_Figure}).
More differential variables are available in Ref.~\cite{ref-cms-ttbb-diff-13TeV}. 


\unskip  

\section{Discussion}

The results from the Run 1 and Run 2 data analyses of ATLAS and CMS of the measured inclusive cross-sections of $\ttbar$ + HF jets are higher than the theoretical predictions (see
Tables~\ref{table:atlas_visible}--\ref{table:full_ratio}).

It will be interesting to observe whether these differences become significant with additional data. 
There were also attempts to measure the differential cross-sections aiming to identify variables where the differences become larger. 
The measured differential cross-sections are in general consistent with theory predictions within its large statistical uncertainty. 
However, in~the $\Delta R$ distribution, there is a discrepancy at the first bin. In~particular, the~\HERWIG prediction tends to produce two additional b jets with smaller angles than the measured value as well as other predictions matched to \PYTHIA. 

In the realm of $\ttbar$ + HF, there is a large fraction of Run-2 data yet to be analyzed and we expect twice more data in Run-3 at the LHC in the coming years. 
We can envisage reducing not only the statistical uncertainty but also systematic uncertainties as more data may enable more data-driven techniques. 
More data will also make it possible to use a smaller bin width to enable hints about potential discrepancies in the inclusive measurements. More advanced heavy-flavor tagging may distinguish the c-flavor jet better from the b-flavor jets. Novel flavor tagging developments can indeed increase our understanding of pQCD and the potential to discover new physics.
More synchronized definitions are also required to compare or combine results from ATLAS and~CMS. 

As we have a systematically higher measured value of cross-section of the $\ttbar$ + HF compared to prediction, more data from Run-3 and eventually from the High Luminosity-LHC may provide interesting opportunities to find cracks in our theoretical understanding. The~discrepancy could be from the fact that signal samples are modeled only at NLO in QCD. We should also make use of the effective field theory (EFT) approach for possible new physics. To~interpret experimental measurements in the context of physics beyond the standard model, the~EFT approach is of interest as a model-independent approach~\cite{EFT_ttbb}. Differential measurements may be crucial in this approach as the presence of the SMEFT operators can modify the kinematics in the standard model~processes. 

\section{Conclusions}

The inclusive and differential cross-sections for the $\ttbar$ + HF jet production have been measured extensively in ATLAS and CMS for the various phase spaces using data samples collected in pp collisions at $\sqrt{s}$ = 7, 8 and 13 TeV. 
The ratio of the cross-sections of the $\ttbar$ + HF jets with respect to the cross-section of the $\ttbar$ + additional jets is also measured, aiming for reduced uncertainties as many kinematic distributions are expected to be similar between the $\ttbar$ + HF jets and the $\ttbar$ + additional jets.  
These measured cross-sections systematically tend to be higher than the predictions.
The measurements are dominated by systematic uncertainties that could be reduced by deploying data-driven techniques to better control the impact of backgrounds and reconstruction-related systematic uncertainties. 
Having more data in the coming years with a better understanding of detectors and more sophisticated reconstruction techniques will bring us to the precision era, where possible new physics could finally be~revealed.

\vspace{6pt} 





\funding{{This work was supported under the framework of the international cooperation program managed by the National Research Foundation of Korea (NRF-2022K2A9A1A06093535). J.D. is supported in part by the Strategic Research Program ``High-Energy Physics'' of the~VUB and also by the FWO-Vlaanderen.}}


\conflictsofinterest{The authors declare no conflict of~interest.} 

\begin{adjustwidth}{-\extralength}{0cm}

\reftitle{References}




\PublishersNote{}
%


\end{adjustwidth}
\end{document}